\documentclass[12pt]{article}

\usepackage{amsmath,amssymb}
\usepackage{graphicx}
\usepackage{epstopdf}
\usepackage{epsfig}
\usepackage{array,multirow}

\newcommand{\gel}{g_{el}}
\newcommand{\gc}{g_*}
\newcommand{\gl}{g}
\newcommand{\Dslash}{\!\not\!\!D}
\newcommand{\ds}{\displaystyle}
\newcommand{\pslash}{\!\not\! p}

\begin{document}


\title{
\vspace*{-1cm}
\hfill{\normalsize\vbox{%
\hbox{\rm \small ROMA1-1441/2006}
}}
\vspace*{3cm} \\
Warped/Composite Phenomenology \\
Simplified \vspace{0.5cm} \\
}

\author{
{
{Roberto Contino$^{\, 1,2}$\thanks{e-mail: roberto.contino@roma1.infn.it}} , 
{Thomas Kramer$^{\, 2}$\thanks{e-mail: tkramer@pha.jhu.edu}} ,
{Minho Son$^{\, 2}$\thanks{e-mail: mhson@pha.jhu.edu}} ,  
{Raman Sundrum$^{\, 2}$\thanks{e-mail: sundrum@pha.jhu.edu}}} \\[0.7cm]
{\small$^1$ \it Dipartimento di Fisica, Universit$\grave{a}$ di Roma ``La Sapienza'' and INFN}\\
{\small \it P.le A. Moro 2, I-00185 Roma, Italy} \\[0.15cm]
{\small$^2$ \it Department of Physics and Astronomy, Johns Hopkins University}\\
{\small \it 3400 North Charles St., Baltimore, MD 21218 USA}
}

\date{}
\maketitle

\begin{abstract}
This is the first of two papers aimed at economically capturing the 
collider phenomenology of warped extra 
dimensions with bulk Standard Model fields, where the 
hierarchy problem is solved non-supersymmetrically.
This scenario is related via the AdS/CFT 
correspondence  to that of partial 
compositeness of the Standard Model. We present a purely 
four-dimensional, two-sector effective field theory describing
the Standard Model fields and just their first 
Kaluza-Klein/composite excitations. This truncation, while losing some 
of the explanatory power and precision of the full higher-dimensional 
warped theory, greatly simplifies phenomenological considerations and 
computations. We describe the philosophy and explicit
construction of our two-sector model, 
and also derive formulas for residual Higgs fine tuning and 
electroweak and flavor precision variables to help identify the most 
motivated parts of the parameter space. We highlight
several of the 
most promising channels for LHC exploration.
The present paper focusses on the most minimal scenario, while the 
companion paper addresses the even 
richer phenomenology of the minimal scenario 
of precision gauge coupling unification. 
\end{abstract}

\newpage


\section{Introduction}

The known non-supersymmetric approaches to the Higgs 
fine-tuning problem
(the Hierarchy Problem) in the Standard Model (SM) are ultimately based on the 
Higgs
degrees of freedom being composite at several TeV, either made from
strongly-coupled constituents or extended objects such as strings.
This is a theoretically challenging arena where the standard tools of
perturbative, renormalizable field theory have limited applicability.
Nonetheless, it is important that experiments intelligently stand watch
for new TeV-scale physics realized in this manner.

In this regard, warped compactifications of higher-dimensional 
spacetime provide a particularly attractive setting, for two reasons:
(i) Their geometry can resolve the Hierarchy Problem 
non-supersymmetrically~\cite{Randall:1999ee},
(ii) When Standard Model fields are
realized in the higher-dimensional ``bulk''~\cite{SMinBulk}, 
quantities of central phenomenological interest become calculable in warped 
effective field theory, thereby describing Kaluza-Klein (KK)
excitations,
flavor hierarchies~\cite{Grossman:2000n, Gherghetta:2000p,Huber:2001sH}, 
rare flavor-violating processes~\cite{Gherghetta:2000p, Huber:2001sH, flavor}, 
electroweak precision 
tests~\cite{RSewpt, Agashe:2003dms, Agashe:2005cp, Agashe:2006c, Agashe:2006crp}, 
gauge-coupling unification~\cite{Unification, Agashe:2005cs}, 
and dark matter~\cite{Agashe:2004s}.
Furthermore, this scenario incorporates or generalizes key features of 
other approaches. Some of these are extra-dimensional mechanisms, such as TeV-scale 
gravity~\cite{TeVstrings, largeED},
Split Fermions~\cite{Arkani-Hamed:2000s}, and the Hosotani 
mechanism~\cite{Hosotani:1983}. But there is also a deep
equivalence or duality, AdS/CFT, between this higher-dimensional 
physics and strongly-coupled, purely 
four-dimensional physics~\cite{AdSCFT1, AdSCFT2, AdSCFT3, AdSCFT4}.
In this way warped compactifications connect closely to the ideas of TeV-scale
strong dynamics, Technicolor~\cite{Weinberg} and 
Walking Technicolor~\cite{WalkingTechnicolor, Csaki:2004gpt}, 
Composite Higgs~\cite{CompositeHiggs1, Contino:2003np}, 
Top-condensation~\cite{TopConden1, Bardeen:1990hl} and 
Topcolor~\cite{Hill:1991}, and partial compositeness~\cite{Kaplan:1991}. 

Calculations in warped effective field theory, while doable, are not 
easy, and they are certainly very difficult to automate by computer.
The purpose of  this paper and its companion,
is to provide a simplifying truncation of the kind of warped
physics (with ``bulk'' SM fields), 
or equivalent composite physics, that could solve the hierarchy 
problem, by
describing only the SM particles and the
first TeV-scale excitations carrying SM charges.~\footnote{The lightest 
SM-neutral graviton and radion excitations of the original Randall-Sundrum 
model \cite{Randall:1999ee} are less central to the phenomenology 
in the present context, but still interesting to consider. 
We have briefly described their inclusion in Appendix E.}
This truncation reproduces
the experimentally accessible new physics to good approximation within 
purely {\it four-dimensional} effective field theory. Calculations are then 
straightforwardly done using Feynman diagrams. 
The price of this truncation is that some of the explanatory power of 
warped compactifications is lost. Enough is retained
to show how existing experimental data and bounds are satisfied, and how
 to identify the least tuned and most promising regions of parameter space.
Roughly, the truncation is achieved by
``deconstruction''~\cite{deconstruction} of the warped theory (discretization of the
warped extra dimension)~\cite{DeconWarp}, making it appear as a variation on Little Higgs
theory~\cite{LH1} (for a review see~\cite{Schmaltz:2005ts}). This
 truncation is very closely related, 
via the AdS/CFT equivalence, to the 
strong interactions approximation known as ``vector meson dominance'', 
in particular as expressed in the formalism of ``hidden local symmetries''
\cite{hidden}. By this route, our work shares some aspects of the
BESS approach \cite{bess} 
to modelling the phenomenology of TeV-scale strong dynamics, with 
however a tighter connection between the new states and the resolution of the
hierarchy problem.
Similarly, our work also shares some features with the Technicolor ``straw man'' 
models of Refs.~\cite{TCSM}.

The model developed in this paper is the minimal one consistent with the 
data as well the central organizing 
principle of {\it partial compositeness}~\cite{Kaplan:1991}. It is closest 
to the TeV-scale physics of the 
most viable and well-developed minimal warped 
models~\cite{Agashe:2003dms, Agashe:2005cp}.
The models of the forthcoming companion paper additionally incorporate
 a simple and striking mechanism for 
precision gauge-coupling unification~\cite{Agashe:2005cs}, with and without a 
weak-scale dark matter candidate~\cite{Agashe:2004s}.
  Our view is that these models reflect grand  
principles, worthy of the significant challenges they pose to experiments, 
and that they are a 
 good point of departure for thinking about how to optimize searches. 
But even if warped compactifications govern TeV physics in Nature,
 it is not guaranteed that any of these models is accurate in 
every detail. Fortunately, the simplicity of their structure  
makes them easy to adapt
in the face of new experimental facts. 

Recently, similar deconstructed approaches to TeV-scale warped phenomenology 
have been 
taken in Refs.~\cite{Chivukula:2006ccshkt, Cheng:2006tw}. 
The present paper has a similar electroweak structure  to that of 
Ref. \cite{Cheng:2006tw}, but differs considerably in other aspects 
such as flavor structure.

In the next section, we will give a broad overview of the physics,
 suppressing technical details, concluding with an outline of the 
remainder of the paper.

\section{Overview}
\label{overview}

\subsection{Two sectors}

Let us  sketch the central physics and how it is captured by the two 
physically equivalent descriptions: four-dimensional strong dynamics, 
and higher-dimensional warped compactification.
We begin with the 
strong dynamics picture, where 
the theory is of the form
\begin{equation}
\label{L}
{\cal L} = {\cal L}_{elementary} + {\cal L}_{composite} + {\cal L}_{mixing}\, . 
\end{equation} 
There is a sector consisting of weakly-coupled 
elementary particles, described by ${\cal L}_{elementary}$. There is a 
second, strongly interacting sector resulting in a host of tightly 
bound composite 
states, including the Higgs doublet, 
described by ${\cal L}_{composite}$. The elementary sector 
couplings are roughly $\gel \sim 1$.
The intra-composite forces holding each composite together are 
very strong, while the residual inter-composite couplings, $\gc$, are 
assumed to be weaker,~\footnote{This is characteristic of 
of gauge theories with large numbers of colors.} 
but still significantly stronger than the elementary couplings, 
$1< \gc \ll 4\pi$. Other than the Higgs boson, the composites are taken to 
have typical masses, $M_*$, of very roughly TeV scale.

\subsection{Partial compositeness}

These two sectors couple to each other via the interactions of 
${\cal L}_{mixing}$, which results primarily in mass-mixing. Consequently,
 mass eigenstates are non-trivial superpositions
 of elementary and composite particles.~\footnote{The SM itself contains 
examples of partial compositeness, where QCD represents the composite 
physics, that may be more familiar to the reader. The SM Higgs vacuum expectation
value (vev) and the  QCD chiral condensate {\it both} break the electroweak symmetry.
As a result, a superposition of QCD-composite pseudoscalars and Higgs 
pseudoscalars (mostly the latter) are eaten by the $W$, $Z$, while the orthogonal 
superposition constitutes the observed light pions. A pion is 
predominantly QCD composite, with a tiny admixture of Higgs pseudoscalar.
Photon-$\rho$ mixing is another example of partial compositeness. 
The fermionic example of 
positron-proton mixing  is forbidden in the SM by its accidental
baryon-number symmetry, but  a small mixing could take place if baryon 
symmetry is broken by non-SM physics such as grand unification.}
The lightest mass eigenstates emerging from Eq.~(\ref{L}) are identified with the SM fields, 
\begin{equation}
\label{SM}
|SM_n \rangle = \cos \varphi_n |{\rm elementary}_n \rangle + \sin 
\varphi_n |{\rm composite}_n \rangle \, ,
\end{equation}
where the mixing angles $\varphi_n$ parametrize the degree of ``partial 
compositeness''. 
The orthogonal admixtures to Eq. (\ref{SM}) constitute the mass eigenstates of the 
TeV-scale new physics. 
When the mass-eigenstate particles interact, the composite components 
interact among themselves with strength $\sim \gc$, and the 
elementary components interact among themselves with 
strength $\sim \gel$. This rough rule captures the essence of 
partially composite phenomenology, which we will further elaborate on.

\subsection{Warped picture}

The warped compactification dual picture of the above physics is that 
the SM and heavy excitations are interpreted as the Kaluza-Klein 
excitations of an extra-dimensional theory, rather than describing them as 
mixtures of elementary and composite degrees of freedom. In this way the 
warped picture more naturally works in terms of the mass eigenstates. 
It is a deep and at first surprising 
result that the two apparently very different theoretical 
descriptions are physically equivalent.
Under favorable circumstances the warped picture can 
give quantitative relations among the most important 
masses, mixing angles and 
 couplings needed for detailed phenomenology. 
By comparison, the composite 
picture   gives a  clear qualitative understanding of many 
issues as well, 
but it is very difficult to pursue phenomenology quantitatively.

\subsection{Truncation}

The truncation of this paper amounts to retaining just the minimal, 
lowest-lying set of composite states needed in the 
SM admixtures.~\footnote{
There is an exception of one charged and one neutral 
additional massive vector mesons,
needed to have an approximate custodial symmetry to protect 
the electroweak $T$ parameter~\cite{Agashe:2003dms}.}  Then,  ${\cal L}_{composite}$
is taken to be a simple effective field theory for these states, while 
${\cal L}_{mixing}$ is taken to be the general set of 
composite-elementary mixing mass terms (compatible with SM gauge invariance).
The mass eigenstates are then the SM states of Eq. (\ref{SM}) and the 
orthogonal heavy states,
\begin{equation}
 |{\rm heavy}_n \rangle = -  \sin \varphi_n |{\rm elementary}_n \rangle + \cos 
\varphi_n |{\rm composite}_n \rangle \, .
\end{equation} 
 This leads to a simple, rough pattern 
of couplings between the SM and heavy states. Let us consider the coupling 
strength for any three mass-eigenstate particles to interact, 
each of which could be a SM particle or a heavy particle. Given the 
basic rule that three elementary components interact with strength 
$\sim \gel$  and three composite components interact with strength $\sim \gc$,
one has
\begin{equation} \label{couplings}
\begin{split}
g_{SM_1 SM_2 SM_3} 
  &\sim \gel \cos \varphi_1 \cos \varphi_2 \cos \varphi_3
   + \gc \sin \varphi_1 \sin \varphi_2 \sin \varphi_3  \\
g_{SM_1 SM_2 {\rm heavy}_3} 
  &\sim - \gel \cos \varphi_1 \cos \varphi_2 \sin 
   \varphi_3 + \gc \sin \varphi_1 \sin \varphi_2 \cos \varphi_3  \\
g_{SM_1 {\rm heavy}_2 {\rm heavy}_3} 
  &\sim  \gel \cos \varphi_1 \sin \varphi_2 \sin 
   \varphi_3 + \gc \sin \varphi_1 \cos \varphi_2 \cos \varphi_3  \\
g_{{\rm heavy}_1 {\rm heavy}_2 {\rm heavy}_3}
  &\sim  - \gel \sin \varphi_1 \sin \varphi_2 \sin \varphi_3
   + \gc \cos \varphi_1 \cos \varphi_2 \cos \varphi_3 \, .
\end{split}
\end{equation}

We now use these results to understand, again roughly, the 
pattern of mixing angles, $\varphi$, needed in the real world, and the 
mechanism that has hidden TeV compositeness from  precision 
tests to date.

\subsection{Compositeness and SM masses}

The SM Higgs doublet is rather special in that it must be
 a full composite, with no elementary 
component, in order to solve the Hierarchy Problem. That is:
 $\sin \varphi_{Higgs}= 1$, $\cos \varphi_{Higgs}=~0$. 
From Eq. (\ref{couplings}) it follows that 
other (pairs of) SM particles couple to it with strength 
$\sim g_* \sin \varphi_1 \sin \varphi_2$,
which determines the extent to which they feel electroweak
symmetry breaking (EWSB), and the masses 
they acquire as a result. From this we deduce that heavier SM 
particles are correlated with larger mixing angles. That is, the heavier SM 
particles have higher partial compositeness, in the operational sense 
that they are more strongly coupled to the new TeV-scale physics. 

In this way the deep question of the origin of the 
 large observed hierarchies in fermion masses translates in the present 
scenario into the question of 
how to obtain large hierarchies among $\varphi_{fermion}$. It is an
 attractive feature of the warped compactification scenario that there is 
a straightforward  mechanism for generating exponential hierarchies of 
just this type. However, this explanatory power is lost 
in our truncation, and the mixing angles, and the necessary hierarchies 
among them, are simply taken as inputs. This is a trade we make in 
favor of simplifying collider phenomenology considerations.

\subsection{Precision tests}

The correlation of SM mass and compositeness helps explain why the virtual 
effects of the new TeV physics have not already been seen in low-energy 
precision tests of the SM, given that electroweak tests are in principle 
sensitive to heavy physics up to $\sim 10$ TeV and flavor-changing tests are 
sensitive to heavy physics up to $\sim 1000$ TeV. 
Experimentally, the
maximum sensitivity applies to the lightest SM particles, but in the present 
scenario the lightest particles, such as  light quarks and leptons, 
have highly $\varphi$-suppressed couplings to the new composite physics. 
For flavor physics, this safety mechanism generalizes the 
Glashow-Iliopoulos-Maiani (GIM) mechanism of 
the minimal SM. In the appendices of this paper  we work out the 
key electroweak precision corrections due to the composite physics as a 
function over the model parameter space, so that due consideration can 
be given to electroweak precision tests. Flavor constraints are 
a bit less decisive due to the freedom of parameters, 
but in Appendix~\ref{FCNC} we illustrate how and to what extent the generalized
GIM mechanism operates within our model.

\subsection{New physics at colliders}

Eq.~(\ref{couplings})  summarizes the challenges as well as
 the  strategy for experimentally searching for the heavy physics. 
Stronger couplings, $g_*$, at first seem promising for production of the 
new physics. But consider the process of fusing 
light SM partons to produce a heavy resonance, 
\begin{equation}
{\cal A}[SM_1 + SM_2 \rightarrow {\rm heavy}] 
 \propto \gc \varphi_1 \varphi_2 \cos \varphi_{heavy} - \gel \sin \varphi_{heavy} \, . 
\end{equation}
The $\gc$ enhancement is (more than) erased by the $\varphi_1 \varphi_2$ 
suppression. This makes it difficult, but not impossible, to produce the 
heavy physics. The challenge will come from seeing these events above 
background. Once produced, these resonances will mostly decay
 to the SM states with the largest mixing angles, namely the heaviest 
SM particles. The efficiency for identifying heavy SM particles, 
in particular the 
top and bottom quarks will be a  key determinant of our ability 
to find the new physics.
But as we will discuss, there are also special opportunities for discovery 
with their own special features, and these merit further study. 

The minimal non-supersymmetric scenario of partial compositeness will be 
 challenging to discover at the Large Hadron Collider (LHC), given the regions of parameter space 
allowed by precision tests. (The scenario with gauge coupling unification, 
discussed in the companion paper to this one has more easily 
accessible physics.) 
This is one of the motivations for the present 
paper: to make the physics as transparent as possible, to motivate 
hard thinking on overcoming experimental obstacles, and to exploit 
special opportunities. 
It is 
particularly important to have a practical measure 
on the parameter space to inform searches of the most motivated regions.
We discuss this next.

\subsection{Residual Higgs fine tuning}

In all known solutions to the Hierarchy Problem, there remains some tension 
between the mechanism for solving the Higgs fine-tuning problem and the 
body of direct searches and precision tests, so that in effect 
one is forced to live with some (more modest) 
residual Higgs fine tuning~\cite{LEPparadox}. This fact is often 
dubbed the ``Little Hierarchy Problem''.  In this paper, we are not 
interested in tracking the residual fine tuning as a means of passing 
judgement on our scenario or in order to debate whether or not it is better
than  others in 
the literature. Rather, we want to use  residual fine tuning  
to determine which regions of our own parameter space are best motivated, 
that is, least tuned. We choose a simple measure of residual fine tuning 
for this purpose,
\begin{equation}
\text{Fine tuning} \sim \frac{m_h^2}{\delta m_{h|mixing}^2}\, .
\end{equation}
The smaller the above ratio, the more tuned a model is. 
We define $\delta m_{h|mixing}^2$ 
to be the largest radiative correction to the {\it physical}
 Higgs mass squared that is 
sensitive to the elementary sector via ${\cal L}_{mixing}$. 
This has to be cancelled (tuned) against the tree-level Higgs mass
to give the  desired $m_h$. 

A useful and elegant feature of our truncated model (and fuller descriptions) 
 is that the leading contributions to this tuning are 
insensitive to the ultraviolet (UV)  cutoff, $\Lambda_{UV}$, and 
unambiguously calculable at one loop. 
Mostly, this result follows by simple power-counting. 
Radiative corrections to the Higgs mass squared that are sensitive to 
the elementary sector must proceed via some virtual fluctuation from 
a composite to an elementary field and back to composite. These fluctuations go through
the mass-mixing interactions of ${\cal L}_{mixing}$, which are at most 
of order of the composite masses, $M_*$, so that 
\begin{equation}
\delta m_{h|mixing}^2 \propto M_*^2 \times {\rm mixing~angles}\, ,
\end{equation}
with at most logarithmic sensitivity to $\Lambda_{UV}$.  
In detail, even this $\log \Lambda_{UV}$ is absent for the dominant 
corrections in our final model. 
In Section~\ref{higgs fine-tuning and finiteness}, the UV-finite fine tuning is simply 
computed as a function over parameter space. 

One may wonder about radiative corrections to the Higgs mass entirely 
originating from the composite dynamics (that is, not going through 
${\cal L}_{mixing}$). In fact, in our truncated model such contributions are 
even larger than $\delta m_{h|mixing}^2$. However, 
we do not count their effect on fine tuning because we expect they would be 
minimized by extra structure in a fuller description of the composite sector. 
The classic example is when the Higgs is realized as a 
composite pseudo-Goldstone boson~\cite{CompositeHiggs1}, such 
as the ordinary pion of QCD, or 
the AdS/CFT dual description in extra dimensions~\cite{Contino:2003np, Agashe:2005cp}, but there 
are other mechanisms as well which deserve further exploration.

\subsection{Full $t_R$ compositeness}

The heaviest SM particle, the top quark, requires sizeable 
mixing angles, $\varphi_{t_L, t_R}$. In fact, since the electroweak 
gauge symmetry implies $\varphi_{t_L} = \varphi_{b_L}$, and the $b_L$ 
has already been sensitively tested, we must conclude that 
this angle is smaller, and, to compensate, $\varphi_{t_R}$ is maximal. 
(A possible exception is if the composite sector has an enlarged custodial
symmetry, a discrete subgroup of which protects the $Zb\bar b$
couplings from receiving large corrections~\cite{Agashe:2006crp}. 
In such case $\varphi_{t_L}$ can be large and $\varphi_{t_R}$ less than maximal.)
Maximal $\varphi_{t_R}$  not only blurs the 
physical distinction between elementary and composite states, 
it also implies that quantum loops involving a virtual $t_R$ 
result in severe Higgs fine tuning. There is a natural way 
out of this difficulty, namely if the $t_R$ is in fact a fully-fledged 
member of the composite sector, without any elementary component. It 
is then 
an integral part of the dynamics that produces 
the light Higgs composite, rather 
than  a large outside destabilizing perturbation. This is reminiscent of 
top-condensate~\cite{TopConden1, Bardeen:1990hl} and 
Topcolor~\cite{Hill:1991} models,
 where the Higgs is realized as a composite of 
strong top quark interactions.

\subsection{The error of our ways}

This paper gives a truncated model describing the SM and the lowest-lying
new excitations, with mass scale $M_*$, with a parametrically 
larger UV cutoff scale, $\Lambda_{UV}$. However, in the full descriptions 
of warped compactifications or composite models the next-to-lowest
excitations typically 
have masses $M_{**} \sim 2 M_*$. Standard effective field theory 
power-counting suggests that the truncated description thereby incurs 
errors of order $(E/M_{**})^2$, where $E$ is the energy of a process under 
consideration. Since we want to consider energies large enough to 
create the first excitations, $E \sim M_*$, we can expect errors of 
about $1/4$. This may seem at first sight rather large, 
  but  the virtual effects of the $M_{**}$-states are
confined to already small corrections to low-energy precision variables or 
small corrections to non-resonant $M_*$-scale cross-sections. These 
corrections are frequently smaller than experimental errors. In practice 
we have found that the truncated model predictions do indeed match well with 
those of related warped compactifications. This makes the 
truncated model a very useful ``reconnaissance tool''
 for our initial experimental forays into the new physics. 

It should be stressed that the truncation procedure we employ for low-lying 
excitations, essentially
discretization or ``deconstruction'' of the warped extra dimension, is part of
a systematically improvable procedure where one makes 
finer and finer discretizations. In 
practice this proliferates the number of sectors in the model (excitation 
levels) and rapidly the continuum description becomes more efficient for 
calculation.

\vspace{1cm}

This concludes our non-technical overview of the relevant physics.
The rest of our paper elaborates in a more quantitative way
on what has been sketched so far, introducing a minimal model 
that gives the simplest realization of partial compositeness
consistent with present data.
Sections \ref{elementary sector} and \ref{composite sector} 
introduce the two halves  of our two-sector model, 
the ``elementary sector'' and ``composite sector'', isolated from 
each other for simplicity. The status of the Higgs doublet as part of the 
composite sector is defined. 
Section~\ref{partial compositeness} introduces the couplings between 
the two sectors in the form of mass-mixing. 
The Lagrangian is then mass-diagonalized 
for phenomenological use. Section~\ref{EWSB} describes the impact of 
electroweak symmetry breaking on 
the masses and couplings of the new physics. 
Section \ref{full tR compositeness} makes the case that 
the $t_R$ should be a composite on par with the Higgs doublet, and shows how 
to simply implement this technically. 
In Section~\ref{higgs fine-tuning and finiteness}, 
it is shown that the dominant radiative corrections to the Higgs mass from 
the elementary sector are UV-finite. These finite corrections are then used to 
estimate Higgs fine tuning as a function of parameter space. 
Section \ref{viable parameter space} 
uses the formulas for the new-physics contributions 
to flavor and electroweak  precision variables computed in
the Appendices \ref{FCNC}, \ref{oblique corrections}
and \ref{sec:ztobbbar}  to
sketch the best motivated regions of the parameter space. 
Section~\ref{pheno} discusses the most promising
channels and strategies to discover the new TeV states at the LHC.
Finally,
Appendix \ref{cutoff} estimates the maximal 
energy to which the model is internally consistent, and shows that this is 
well above upcoming experimental reach. Appendix E briefly describes the 
inclusion within the composite sector 
of the SM-neutral graviton and radion excitations of the 
Randall-Sundrum model.

\section{Elementary Sector}
\label{elementary sector}

In this section we define the first building block of
our minimal two-site model of partial compositeness: the
elementary sector.
Its field content corresponds precisely to that of the 
SM, but with the notable exception of the Higgs field. 
Indeed, the elementary fields will ultimately 
constitute the dominant component of the physical SM fermions and gauge 
bosons. The elementary gauge fields, corresponding to adjoints of
the elementary gauge group
$\left [SU(3)_c \otimes SU(2)_L \otimes U(1)_Y \right ]^{el}$, 
are denoted by
\begin{equation}\label{ele gauge fields}
A_{\mu} \equiv \left\{ G_{\mu}, W_{\mu}, {\cal B}_{\mu} \right\} \, .
\end{equation}
The SM electroweak doublet fermions are denoted by
\begin{equation}
\label{L ele fermion}
\psi_L \equiv \left\{ q_{Li} = 
(u_{Li}, d_{Li}), \ell_{Li} = (\nu_{Li}, e_{Li}) \right\}\, , 
\qquad i = 1,2,3 \, ,
\end{equation}
or by their more individual names such as $t_L, \nu_{eL}, \mu_L$.
The SM electroweak singlet fermions are denoted by~\footnote{The 
collective symbol $\tilde{\psi}_R$ is deliberately redundant. Normally, 
right-handed SM fermions are assumed to be electroweak singlets. Here, we 
additionally use the tilde to denote this fact and conform with our notation 
for the composite sector, where both chiralities of electroweak singlets 
ultimately appear.}
\begin{equation}
\label{R ele fermion}
\tilde{\psi}_R \equiv  \left\{ u_{Ri}, d_{Ri}, \nu_{Ri}, e_{Ri} \right\}\, .
\end{equation}

The only renormalizable interactions in this sector are gauge interactions,  
\begin{equation}
\label{ele action}
{\cal L}_{elementary} 
= - \frac{1}{4} F_{\mu \nu}^2 + \bar{\psi}_L i \Dslash \psi_L
+ \bar{\tilde{\psi}}_R i \Dslash \tilde{\psi}_R \, .
\end{equation}
The associated gauge couplings, $g_{el\, 1}$, $g_{el\, 2}$, $g_{el\, 3}$,
will turn out to be approximately, but not exactly, 
equal to the measured SM gauge couplings, $g_{1}$, $g_{2}$, $g_{3}$.
For technical reasons discussed below, it is more convenient to use the 
standard SO(10) grand unified theory (GUT) convention for the hypercharge 
\begin{equation} \label{hypercharge}
g_{1} \equiv \sqrt{\frac{5}{3}}\, g_Y,\, \qquad Y_{GUT}\equiv\sqrt{\frac{3}{5}}\,Y \, ,
\end{equation}
where $g_Y$ and $Y$ denote the hypercharge coupling and hypercharge generator
in the usual SM convention, and $Y_{GUT}$ is defined by Eq.~(\ref{hypercharge}).

The elementary sector gauge dynamics obviously
makes sense in isolation to the highest scales, say the 
Planck scale. 
Non-renormalizable operators made of elementary fields,
like for example flavor and CP violating four-fermion  
interactions, will be strongly suppressed and thus negligible.
Any effect of flavor and CP violation must therefore come from the
composite sector and proceed through its interactions with the 
elementary fields.

\section{Composite Sector}
\label{composite sector}

The composite sector comprises the Higgs plus what is 
essentially an ``excited''  
copy of the SM fermions and gauge bosons. The basic role of these 
excitations is 
to provide a small composite component to the physical SM fermions and 
gauge bosons, which determines the extent to which the latter couple to the 
Higgs and eventually feel  electroweak symmetry breaking. 
This would naively suggest, in particular, that the massive vector  
excitations $\rho_{\mu}$ should correspond to an adjoint of $SU(3) \otimes 
SU(2) \otimes U(1)$.
However, in order to protect the electroweak $T$ parameter as discussed below, 
the composite bosons must respect a larger symmetry, minimally
$\left [ SU(3) \otimes SU(2) \otimes SU(2) \otimes U(1) \right ]$.
In a controlled effective field theory, massive vector mesons must always be 
realized as gauge bosons of a broken gauge group.
We therefore take the  $\rho_{\mu}$ as 
gauge bosons (hence adjoint representations) of a group 
$\left [ SU(3)_c \otimes SU(2)_L \otimes SU(2)_R \otimes U(1)_X \right ]^{comp}$. 
The SM Higgs field is assumed to transform as a real bidoublet $(\tilde{H},H)$
under $\left[SU(2)_L\otimes SU(2)_R\right]^{comp}$, while the quantum numbers
that we adopt for the massive fermion excitations, $\chi$, $\tilde{\chi}$, are 
those given in Table~\ref{tab:qnumb}.~\footnote{The factor of $\sqrt{3/2}$ 
in the $U(1)_X$ charges comes from the relation
$Y_{GUT}= \left(T^{3R} + \xi \, T_X\right)/\sqrt{1+ \xi^2}$,
where $\xi = \sqrt{2/3}$ and $T^{3R}=0$ for all the fermions.
See sec.~\ref{vector mesons} for more details.}
\begin{table}[t]
\begin{center}
\setlength\extrarowheight{5pt}
\begin{tabular}{|c|c|c|c|c|c|} 
\hline
\multicolumn{2}{|c|}{} & $SU(3)_c$ & $SU(2)_L$ & $SU(2)_R$ & $U(1)_X$ \\[5pt] 
\hline\hline 
\multicolumn{2}{|c|}{$\rho_\mu$} &\multicolumn{4}{c|}{Gauge Fields}\\[5pt]
\hline
\multicolumn{2}{|c|}{$(\tilde H, H)$} & $\bf1$ & $\bf2$ & $\bf2$ & 0 \\[5pt]
\hline
\multirow{2}{2.5mm}{$\chi$} & Q & $\bf3$ & $\bf2$ & $\bf1$ & $\frac{1}{6} \cdot \sqrt{\frac{3}{2}}$ \\[5pt]
\cline{3-6}
&L & $\bf1$ & $\bf2$ & $\bf1$ & $\left(-\frac{1}{2}\right) \cdot \sqrt{\frac{3}{2}}$  \\[5pt]
\hline
\multirow{5}{2.5mm}{$\tilde\chi$} &$\tilde U$ & $\bf3$ & $\bf1$ & $\bf1$ 
                          & $\frac{2}{3} \cdot \sqrt{\frac{3}{2}}$  \\[5pt]
\cline{3-6}
&$\tilde D$ & $\bf3$ & $\bf1$ & $\bf1$ 
                          & $\left(-\frac{1}{3}\right) \cdot \sqrt{\frac{3}{2}}$  \\[5pt]
\cline{2-6}
&$\tilde N$ & $\bf1$ & $\bf1$ & $\bf1$ & 0 \\[5pt]
\cline{3-6}
&$\tilde E$ & $\bf1$ & $\bf1$ & $\bf1$ & $\left(-1\right) \cdot \sqrt{\frac{3}{2}}$  \\[5pt]
\hline
\end{tabular}
\end{center}
\caption{\it Field content and quantum numbers
in the composite sector of the two-site minimal model.}
\label{tab:qnumb}
\end{table}

The composite dynamics is then summarized by
\begin{equation}
\begin{split}
\label{comp action}
 {\cal L}_{composite} = 
& - \frac{1}{4} \rho_{\mu \nu}^2 + \frac{M_*^2}{2} \rho_{\mu}^2 +
  |D_{\mu} H|^2 - V(H) \\
& + \bar{\chi} (i \Dslash - m) \chi
  + \bar{\tilde{\chi}} (i \Dslash - \tilde{m}) \tilde{\chi}
  - \bar\chi \big( Y_{*u} \tilde H \tilde\chi^u + Y_{*d} H \tilde\chi^d \big) + \text{h.c.} \, ,
\end{split}
\end{equation}
where $\tilde\chi^u \equiv \{ \tilde U , \tilde N \}$, 
$\tilde\chi^d \equiv \{ \tilde D , \tilde E \}$.
The $\rho_{\mu}$ mass terms clearly break the composite gauge invariance completely.
Technically, one can imagine that 
this is due to a Higgs mechanism 
 (distinct from the electroweak Higgs mechanism) in which the 
associated Goldstone bosons have been eaten by the massive $\rho$'s (that 
is, we are in unitary gauge), and other related 
massive fluctuations are omitted because their masses are at the 
cutoff scale of our composite effective field theory description. 
The perturbativity of our
effective description then implies an upper bound on this cutoff, which we 
estimate in Appendix~\ref{cutoff}. 
It can be large enough  so that 
unspecified physics at this scale decouples from precision constraints on the
model as well as collider search considerations. 
Closer inspection shows that the Yukawa couplings 
further break the gauge invariance,
and this is also taken into account in Appendix~\ref{cutoff} 
in estimating the maximal cutoff.

The mass parameters and couplings of ${\cal L}_{composite}$ are chosen so that 
the purely bosonic subsector has an
$\left [ SU(3)_c \otimes SU(2)_L \otimes SU(2)_R \otimes U(1)_X \right 
]$ \it global \rm symmetry, while the Yukawa couplings break this down to 
an $\left [ SU(3)_c \otimes SU(2)_L  \otimes U(1) \right ]$ 
global symmetry. As mentioned above, the higher bosonic symmetry will 
be important in protecting the electroweak $T$ parameter. 
It might at first seem unnatural to have a subsector
of a theory enjoy a higher symmetry than the full theory, but in the present 
context it is technically natural because of the following 
observation. The Yukawa couplings do break the $SU(2)_R$ symmetry, 
group-theoretically by $\Delta I_R = 1/2$. But  
($SU(2)_R$)-breaking in $M_*^2$ (while preserving the total 
$\left [ SU(3)_c \otimes SU(2)_L  \otimes U(1) \right 
]$ symmetry) is necessarily $\Delta I_R = 2$. Thus, 
quadratically divergent 
($SU(2)_R$)-breaking radiative corrections to $M_*^2$ due to the Yukawa 
couplings are proportional to $Y_*^4$
and arise at three (composite) loops.
With the cutoff estimates of Appendix~\ref{cutoff} and composite couplings 
large but still in the perturbative range, $\gc , Y_* \ll 4\pi$,
this breaking is negligible for our purposes. 
Finally, the Lagrangian ${\cal L}_{composite}$ is assumed to have  
a parity symmetry for simplicity, which means equal Yukawa
couplings for both chiralities of the massive fermions.

Having sketched the composite dynamics and symmetry realizations, we will 
now discuss the  fields and couplings in greater 
detail.

\subsection{Vector mesons}
\label{vector mesons}

In principle, the couplings $g_*$ and
the masses $M_*$ of the vector mesons could be distinct for each 
simple subgroup of $\left [ SU(3)_c \otimes SU(2)_L \otimes
 SU(2)_R \otimes U(1)_X \right ]^{comp}$.  In the present paper we will 
assume, purely for convenience, that these couplings are the same for the 
$SU(2)_R \otimes U(1)_{X}$ subgroup of the composite gauge group. 
We thereby assign the parameters $\left(g_{*}  , M_{*}\right)_{1}$, 
$\left(g_{*}  , M_{*}\right)_{2}$, and $\left(g_{*}  , M_{*}\right)_{3}$ to the
$SU(2)_R \otimes U(1)_X$, $SU(2)_L$, and $SU(3)_c$ composite gauge groups.  
As a further simplification in subsequent numerical estimates and calculations,
we will assume that all the $M_{*\, i}$ are similar, and that all the composite
couplings are similar and moderately large, roughly $2 < g_{*\, i} < 4$.

We  subdivide the 
vector mesons into those that will ultimately mix with the elementary 
gauge bosons, $\rho^*_\mu$,
and those that will not, $\tilde\rho_\mu$:
\begin{equation}
\label{entire vector meson}
\rho_\mu = \{ \rho^*_\mu, \tilde\rho_\mu \}\, .
\end{equation}
The $\rho^*_\mu$ transform as an adjoint
of the exact global symmetry 
$\left [ SU(3) \otimes SU(2)_L \otimes U(1)_Y \right ]$,
while the $\tilde\rho_\mu$ form an orthogonal combination:~\footnote{
Our notation for the excited hypercharge boson is chosen so as not to clash 
with that of the excited bottom quark discussed in Section~\ref{composite fermions}.}
\begin{equation}
\label{rho mesons}
\rho^*_\mu = \{ G_\mu^* , W_\mu^* ,{\cal B}_\mu^* \}\, , \qquad
 \tilde\rho_\mu = \Big\{ \tilde W^{\pm}_\mu \equiv \frac{\tilde W_1 \mp i\, \tilde W_2}{\sqrt{2}} , 
 \tilde {\cal B}_\mu \Big\} \, .
\end{equation}
We associate the generators $T^{1R},T^{2R}$ to the fields $\tilde W^{1},\tilde W^2$
and $T_{\cal{B}^* }\equiv Y_{GUT}=(T^{3R}+\xi\, T_X )/\sqrt{1+\xi^2}$, 
$T_{\tilde {\cal B}}=(\xi\, T^{3R}-\, T_X )/\sqrt{1+\xi^2}$
respectively to $\cal{B}^* $ and $\tilde {\cal B}$.
We choose $\xi = \sqrt{2/3}$ to match the SO(10) GUT normalization. In this way the
generator of the exactly preserved \it global \rm $U(1)$ is just the hypercharge in SO(10) 
GUT normalization (i.e. $Y_{GUT}=\sqrt{3/5}\,Y$).
This choice of normalization also allows us to have symmetric expressions for all gauge bosons 
after mass diagonalization as seen in Section \ref{partial compositeness}.

\subsection{The Higgs field}
\label{the higgs field}

The SM Higgs field is assumed to be entirely a composite of the 
new strong dynamics at TeV scales. Its bi-doublet transformations under 
$[SU(2)_L \otimes SU(2)_R]^{comp}$ are simply written by expressing the Higgs 
in $2 \times 2$ matrix form,
\begin{equation}
(\tilde H,H) \rightarrow L \, (\tilde H,H) \, R^\dagger \, ,
\end{equation}
where $H$ is the usual SM Higgs doublet field for the down-type Yukawa couplings and 
$\tilde H = i\sigma^2 H^*$ for the up-type Yukawa couplings.
The classical Higgs potential has the  usual SM form,
\begin{equation}
V(H) \equiv - \mu_H^2 |H|^2 + \lambda_H |H|^4 \, ,
\end{equation}
and is $[SU(2)_L \otimes SU(2)_R]^{comp}$-invariant.

A critical  assumption we will make is that 
a fuller description of the composite
dynamics (neglecting other sectors such as the elementary sector)  
would make $V(H) \equiv 0$ natural (for example, if the Higgs were 
realized as a Goldstone boson) and that there may in addition be 
perturbing dynamics in the composite sector 
that yield a non-vanishing but weak potential, 
\begin{equation}
\mu_H  < {\rm TeV}\, ,  \qquad
\lambda_H \sim 1 \, .
\end{equation}
That is we will use this simple classical $V(H)$ to stand for the fully 
renormalized effective potential due to the composite dynamics alone, 
and assume that the desired values of $\mu_H, \lambda_H$ require no 
tuning {\it before radiative corrections from the elementary sector is taken 
into account}. Those  radiative corrections 
to the Higgs potential that are external to the composite sector
will determine our measure of fine tuning.

\subsection{Composite fermions}
\label{composite fermions}

The composite fermions parallel the SM fermions closely, except for the 
fact that they are massive Dirac fermions, rather than Weyl fermions. 
There are three generations of 
$SU(2)_L$ doublet composite Dirac fermions, denoted by 
\begin{equation}
\label{su2L comp fermion}
\chi \equiv \{ Q_{i} = (U_i, D_i), L_i = (N_i, E_i)\} \, ,
\end{equation}
or by more familiar names such as $(T, B), N_{\mu}, E_{\tau}$.
The $SU(2)_L$ singlet composite Dirac fermions are denoted by
\begin{equation}
\label{su2R comp fermions}
\tilde{\chi} \equiv \{ \tilde{U}_i, \tilde{D}_i, \tilde{N}_{i}, \tilde{E}_{i} \} \, ,
\end{equation}
or by even more familiar names such as $\tilde{T}, \tilde{B}, \tilde{N}_{\mu}, \tilde{E}_{\tau}$.
Since the composite fermions have $SU(2)_L$  doublets and singlets with both 
left- and right-handed Lorentz representations, we use the 
presence of a tilde to denote the singlets, while we use subscripts ``$L, R$'' to 
denote the Lorentz representation. So, for example, ``$\tilde{C}_L$'', 
denotes the left-handed Lorentz chirality of the $SU(2)_L$-singlet
``charmed'' composite quark.

The Yukawa couplings between $SU(2)_L$ singlets and doublets are denoted by
\begin{equation}\label{collective yukawa couplings}
 Y_{*u} = \{Y_{*U},Y_{*N} \},\, \qquad 
 Y_{*d} = \{Y_{*D},Y_{*E} \} \, .
\end{equation}
We are assuming for simplicity that ${\cal L}_{composite}$ 
has a parity symmetry, which means equal Yukawa
couplings for both chiralities of the massive fermions.
Here $Y_{*u}, Y_{*d}$
are matrices in ordinary generational space, 
which we also take to be stronger and less hierarchical than their 
SM equivalents, $1 < Y_{*u,d} \lesssim 3$,  again a reflection of the  
strong dynamics creating the composites. 
The Dirac mass terms, $m, \tilde{m}$ are free parameters of the model. 
They respect the $\left [ SU(3)_c \otimes SU(2)_L \otimes SU(2)_R \otimes U(1)_X \right ]$
global symmetry of the composite sector, and are independent for each 
representation of
massive fermions.
Purely for technical simplicity, we further require these mass terms to be 
predominantly diagonal $3 \times 3$  matrices in generation space. 
That is, we are taking the 
composite Yukawa couplings to be the dominant source of breaking 
of generational $U(1)$ symmetries. Subdominantly, radiative corrections from 
Yukawa couplings will necessarily 
make the mass terms off-diagonal, but this will not play an important role.

\section{Partial Compositeness}
\label{partial compositeness}

Partial compositeness is realized by adding a set of mass-mixing
(soft mixing) terms to our model, 
\begin{equation} \label{soft mix}
{\cal L}_{mixing} =
 - M_*^2\, \frac{\gel}{\gc}\, A_{\mu} \rho_{\mu}^* + 
   \frac{M_*^2}{2} \left( \frac{\gel}{\gc} A_{\mu} \right)^2
   + (\bar\psi_L \Delta \chi_R + \bar{\tilde\psi}_R \tilde\Delta \tilde\chi_{L} + {\rm h.c.})\, ,
\end{equation}
where an implicit sum over all species of gauge and fermionic fields is understood.

\subsection{Residual Standard Model Gauge Invariance}

The vector boson terms reflect the 
 gauging of the global symmetry group of the composite sector of Section~\ref{composite sector} 
 by the elementary gauge symmetry. This looks somewhat unfamiliar because 
the (broken) $\rho$ gauge symmetry is being treated in unitary gauge.
Indeed, it is straightforward to check that the entire Lagrangian, Eq.~(\ref{L}), 
is exactly invariant under an $SU(3)\otimes SU(2) \otimes U(1)$
{\it gauge symmetry} which we identify with the final SM gauge symmetry. The 
corresponding SM gauge fields are then superpositions 
\begin{equation}\label{SM gauge boson}
\frac{\gc}{\sqrt{g_{el}^2 + g_*^2}}\, A_{\mu} + 
\frac{\gel}{\sqrt{g_{el}^2 + g_*^2}}\, \rho^*_{\mu}\, ,
\end{equation}
and the SM gauge couplings have the form
\begin{equation}\label{SM gauge coupling}
g = \frac{g_{el}\, g_*}{\sqrt{g_{el}^2 + g_*^2}} \simeq g_{el}\, , \quad 
 \text{for }  g_{el} \ll g_* \, .
\end{equation}
The non-trivial superposition of  $A_{\mu}$ and $\rho^*_{\mu}$ inside the 
SM gauge field is the vector meson version of partial compositeness. 
In the literature of low-energy hadronic phenomenology 
the analogous phenomenon is known as ``photon-$\rho$'' mixing.

The fermionic terms in Eq.~(\ref{soft mix}) describe fermionic partial compositeness in 
terms of mixing-mass parameters $\Delta$, $\tilde{\Delta}$, which are 
independent for each fermionic SM gauge-representation (or, equivalently,
for each species of massive fermion).
In the interest of simplicity, they are chosen to be diagonal in the same basis as the 
$m$ and $\tilde{m}$ are, in generation space.
They explicitly break the separate elementary and composite
gauge symmetries of the $\rho_{\mu}$ and $A_{\mu}$ but 
preserve the SM gauge invariance discussed above. 
The mass-mixing of  elementary chiral fermions with 
composite Dirac fermions necessarily results in a new set of 
massless chiral fermions, which are linear combinations of the original 
$\chi$ and $\psi$, and are identified with the SM fermions.

\subsection{Mass Eigenstates}
Most phenomenological aspects of the model related to the 
production and detection of the new massive particles at the
colliders are better pursued in the 
canonical language of diagonalized mass and kinetic terms. 
We diagonalize the mass mixing 
arising from ${\cal L}_{mixing}$ by field transformations:
\begin{align}
\label{gauge mass diag}
\begin{pmatrix} A_\mu \\ \rho^*_\mu \end{pmatrix}  &\rightarrow 
 \begin{pmatrix} \cos\theta & -\sin\theta \\ \sin\theta & \cos\theta \end{pmatrix}
 \begin{pmatrix} A_\mu \\ \rho^*_\mu \end{pmatrix}\, , \quad &
 \tan\theta &= \frac{\gel}{\gc}\, , \\[0.3cm]
\label{L fermion mass diag}
\begin{pmatrix} \psi_L \\ \chi_L \end{pmatrix}  &\rightarrow 
 \begin{pmatrix} \cos\varphi_{\psi_L} & -\sin\varphi_{\psi_L} \\ 
   \sin\varphi_{\psi_L} & \cos\varphi_{\psi_L} \end{pmatrix}
 \begin{pmatrix} \psi_L \\ \chi_L \end{pmatrix}\, , \quad &
 \tan\varphi _{\psi_L} &= \frac{\Delta}{m}\, , \\[0.3cm]
\label{R fermion mass diag}
\begin{pmatrix} \tilde\psi_{R} \\ \tilde\chi_{R} \end{pmatrix}  &\rightarrow 
 \begin{pmatrix} \cos\varphi_{\tilde\psi_{R}} & -\sin\varphi_{\tilde\psi_{R}} \\ 
   \sin\varphi_{\tilde\psi_{R}} & \cos\varphi_{\tilde\psi_{R}} \end{pmatrix}
 \begin{pmatrix} \tilde\psi_{R} \\ \tilde\chi_{R} \end{pmatrix}\, , \quad &
 \tan\varphi_{\tilde\psi_{R}} &= \frac{\tilde\Delta}{\tilde m}\, .
\end{align}
Our notation above has been chosen to be economical with symbols. Before 
the diagonalization, ``$A_{\mu}, \psi_L, \tilde\psi_R$'' denoted 
the elementary fields and ``$\rho^*_{\mu}, \tilde{\rho}_{\mu}, 
\chi, \tilde{\chi}$'' the composite fields. After the diagonalization, 
``$A_{\mu}, \psi_L, \tilde\psi_R$'' denote the SM fields, which are 
massless before EWSB, while 
``$\rho^*_{\mu}, \tilde{\rho}_{\mu}, 
\chi, \tilde{\chi}$'' denote new-physics mass eigenstates (before EWSB).

The mixing angles relating the elementary/composite basis to the mass 
basis (before EWSB) now parametrize partial compositeness.
Note that all the mixing angles are real and there is one for every 
SM multiplet, 
\begin{equation}
\label{collectivesymbols}
\begin{split}
\theta &\equiv \theta_1, \theta_2, \theta_3 \\
\varphi_{\psi_L} &\equiv \varphi_{q_{Li}}, \varphi_{\ell_{Li}} \\
\varphi_{\tilde\psi_{R}} &\equiv \varphi_{u_{Ri}}, \varphi_{d_{Ri}}, 
\varphi_{\nu_{Ri}} , \varphi_{e_{Ri}}\, .
\end{split}
\end{equation}
The indices for $\theta$ refer to the $SU(3) \otimes SU(2) \otimes U(1)$
factors of the SM gauge group, while the `$i$' indices on the 
$\varphi = \{ \varphi_{\psi_L}, \varphi_{\tilde\psi_{R}} \}$ are 
generational.

\subsection{Diagonalized Lagrangian}

Implementing the transformation above in the total Lagrangian,
\begin{equation}
\label{entire exp action}
{\cal L} = {\cal L}_{gauge} + {\cal L}_{fermion} + {\cal L}_{Higgs} \, , 
\end{equation}
gives:
\begin{align}
\label{exp gauge boson action}
\begin{split}
{\cal L}_{gauge} =  
 &- \frac{1}{4} F_{\mu \nu}^2 \\
 &+ \frac{1}{2} \left( D_\mu\rho_\nu D_\nu\rho_\mu -  D_\mu\rho_\nu D_\mu\rho_\nu \right) 
  + \frac{M_{*1}^2}{2}\, \tilde\rho_\mu^2 +  \frac{M_*^2}{2 \cos^2\theta}\, \rho_\mu^{*\, 2} 
  + \frac{i\gl}{2} F_{\mu\nu} [\rho_\mu,\rho_\nu] \\
 &+ 2 i \gl \cot 2\theta\, D_\mu\rho^*_\nu [ \rho^*_\mu,\rho^*_\nu ]
  + \frac{i \gl_1}{\sin\theta_1}\, D_\mu\tilde\rho_\nu [ \tilde\rho_\mu,\tilde\rho_\nu ] \\
 &+ i \gl_1\cot\theta_1\, D_\mu\rho^*_\nu [\tilde\rho_\mu,\tilde\rho_\nu ]
  + i \gl_1\cot\theta_1\, D_\mu\tilde\rho_\nu \left([\rho^*_\mu,\tilde\rho_\nu] 
  + [\tilde\rho_\mu,\rho^*_\nu]\right) \\
 &+\frac{\gl^2}{4} \left( \frac{\sin^4\theta}{\cos^2\theta} + \frac{\cos^4\theta}{\sin^2\theta} \right)
   [\rho_\mu^*,\rho^*_\nu]^2 + \frac{\gl_1^2}{4\sin^2\theta_1}\,   [\tilde\rho_\mu,\tilde\rho_\nu]^2 \\
 &+\frac{\gl_1^2}{4}\cot^2\theta_1 \left( [\rho^*_\mu,\tilde\rho_\nu] + [\tilde\rho_\mu,\rho^*_\nu] \right)^2
  + \gl_1^2 \frac{\cos\theta_1}{\sin^2 \theta_1} [\tilde\rho_\mu , \tilde\rho_\nu][\rho^*_\mu,\tilde\rho_\nu]\, ,
\end{split} \\[0.5cm]
\label{exp fermion action}
\begin{split}
{\cal L}_{fermion} =
 & \bar{\psi}_L i \Dslash \psi_L + \bar{\chi} (i \Dslash -  m_*) \chi \\
 &+ \bar\psi_L \!\left[ \gl \left( \sin^2\varphi_{\psi_L} \cot\theta - \cos^2\varphi_{\psi_L} \tan\theta \right) 
     \rho^*_\mu + \frac{\gl_1}{\sin\theta_1}\sin^2\varphi_{\psi_L}\, \tilde\rho_\mu \right] \gamma^\mu \psi_L \\
 &+ \bar\psi_L \left( \gl
    \frac{\sin\varphi_{\psi_L} \cos\varphi_{\psi_L}}{\sin\theta \cos\theta}\, \rho^*_\mu +
    \frac{\gl_1}{\sin\theta_1}\sin\varphi_{\psi_L} \cos\varphi_{\psi_L}\, \tilde\rho_\mu \right) \gamma^\mu \chi_L + {\rm h.c.} \\
 &+ \bar\chi_L \!\left[ \gl\left( \cos^2\varphi_{\psi_L} \cot\theta - \sin^2\varphi_{\psi_L} \tan\theta \right) 
     \rho^*_\mu + \frac{\gl_1}{\sin\theta_1}\cos^2\varphi_{\psi_L}\, \tilde\rho_\mu \right] \gamma^\mu \chi_L \\
 &+ \bar\chi_R \left( \gl\cot\theta\, \rho^*_\mu + \frac{\gl_1}{\sin\theta_1}\, \tilde\rho_\mu \right) \gamma^\mu \chi_R \\
 &+ \{ L \leftrightarrow R\, ; \, \chi \rightarrow \tilde\chi \, ; \, 
       \varphi_{\psi_L} \rightarrow \varphi_{\tilde\psi_{R}} \, ; \, m_* \rightarrow \tilde m_* \}\, ,
\end{split}
\end{align}
%
%
\begin{equation}
\label{exp higgs action}
\begin{split}
{\cal L}_{Higgs} =
 & |D_\mu H|^2 - V(H) \\
 & + H^\dagger i\, \gl\cot\theta\, \rho^*_\mu D_\mu H + {\rm h.c.}\\
 & - i \frac{\gl_1}{2\sin\theta_1}\, \left (\frac{1}{\sqrt{2}}\, \tilde H^\dagger \tilde W^-_\mu D_\mu H 
   + \frac{1}{\sqrt{2}}\, H^\dagger \tilde W^+_\mu D_\mu \tilde H - \sqrt{\frac{2}{5}} H^\dagger \tilde {\cal B}_\mu D_\mu H \right ) 
   + {\rm h.c.} \\
 &- \gl_1 \gl \frac{\cot\theta}{\sin\theta_1}
  \left ( \frac{1}{\sqrt{2}}\, \tilde H^\dagger \rho^*_\mu  \tilde W^-_\mu H
  + \frac{1}{\sqrt{2}}\, H^\dagger \rho^*_\mu \tilde W^+_\mu \tilde H
  - \sqrt{\frac{2}{5}} H^\dagger \rho^*_\mu \tilde {\cal B}_\mu H  \right ) \\
 & + H^\dagger \left \{ \left ( \gl\cot\theta\, \rho^*_\mu \right )^2 
   + \frac{\gl^2_1}{\sin^2\theta_1}\left (\frac{1}{2}\tilde W^+_\mu \tilde W^-_\mu 
   + \frac{1}{10} \tilde {\cal B}^2_\mu \right ) \right \} H \\
 &- \left( \sin\varphi_{\psi_L}\, \bar\psi_L + \cos\varphi_{\psi_L}\, \bar\chi_L \right)  
    \Big[ Y_{*u} \tilde H \!
    \left( \sin\varphi_{\tilde\psi^u_{R}}\, \tilde\psi^u_{R} + \cos\varphi_{\tilde\psi^u_{R}}\, \tilde\chi^u_{R} \right) \\
  & \hspace{5.3cm} +  Y_{*d} H \!
    \left( \sin\varphi_{\tilde\psi^d_{R}}\, \tilde\psi^d_{R} + \cos\varphi_{\tilde\psi^d_{R}}\, \tilde\chi^d_{R} \right) 
    \Big] + {\rm h.c.} \\
 &-\bar\chi_R Y_{*u} \tilde H \tilde\chi^u_{L} -\bar\chi_R  Y_{*d} H \tilde\chi^d_{L} + {\rm h.c.} \, ,
\end{split}
\end{equation}
where $\tilde\psi^u_R \equiv \{ u_R , \nu_R\}$, $\tilde\psi^d_R \equiv \{ d_R , e_R\}$, and, we recall,
$\tilde\chi^u \equiv \{ \tilde U , \tilde N \}$, $\tilde\chi^d \equiv \{ \tilde D , \tilde E \}$.
In Eqs.~(\ref{exp gauge boson action}), (\ref{exp fermion action})
and (\ref{exp higgs action}), all covariant derivatives of fermions and heavy gauge bosons
are now with respect to the unbroken SM gauge group, 
\begin{equation}\label{exp covariant derivative}
D_{\mu} \equiv \partial_{\mu} - i \gl A_{\mu} \, , \qquad \gl =  \gc \sin\theta \, ,
\end{equation}
and
\begin{equation}
\label{exp mass}
m_* \equiv \sqrt{\Delta^2 + m^2}\, , \qquad 
 \tilde m_{*} \equiv \sqrt{\tilde\Delta^2 + \tilde m^2} \, .
\end{equation}
All vector fields in Eqs.~(\ref{exp fermion action}) and (\ref{exp higgs action}), 
(including those in the covariant derivatives) are to be considered in a matrix notation, 
i.e. each gauge component multiplies its corresponding generator $T^a$, normalized according to the
standard convention ($\text{Tr} (T^a T^b) = \delta^{ab}/2$ for the non-abelian generators).
The only exception is for $\tilde W^{\pm}$ and $\tilde {\cal B}$ 
in Eq.~(\ref{exp higgs action}), which are component fields.
This complication arises as a result of the subtle transformation of the Higgs under 
$\left[SU(2)_R\right]^{comp}$.
In the gauge Lagrangian (\ref{exp gauge boson action}) we adopt a different notation:
gauge fields are still in matrix notation, with an implicit trace operation over the whole Lagrangian,
but with the following normalization: $\text{Tr} (T^a T^b) = \delta^{ab}$ for non-abelian generators, 
$T=1$ for the abelian ones. This choice leads to a more compact and simple-to-read expression, compared
to the standard normalization or the use of component fields.
Gauge couplings, mixing angles and masses of the heavy gauge bosons 
without explicit indices must be 
understood as collective symbols (see Eq.~(\ref{collectivesymbols}) 
as well):~\footnote{As declared in Section~\ref{vector mesons}, 
we are assuming for simplicity a common mass parameter $M_{*1}$ for 
$\tilde W^\pm$, $\tilde {\cal B}$, and~${\cal B}^*$.}
\begin{equation}
 \gl = \{\gl_3, \gl_2, \gl_1 \}, \qquad M_* = \{M_{*3},M_{*2},M_{*1} \}\, ,
\end{equation}
and similarly for the masses and Yukawa couplings of the heavy fermions
(see also Eq.~(21)):
\begin{equation} \label{heavy masses}
m_* \equiv \{ m_*^Q , m_*^L \} \, , \qquad 
\tilde m_* \equiv \{ \tilde m_*^U , \tilde m_*^D , \tilde m_*^N , \tilde m_*^E \} \, .
\end{equation}
With the definitions of the collective symbols of the fields, Eqs.~(\ref{rho mesons}),
(\ref{su2L comp fermion}) and (\ref{su2R comp fermions}), 
all terms in the action after mass
diagonalization can be easily decoded by summing over the implicit indices.

The effective field theory (EFT) of 
Eqs.~(\ref{exp gauge boson action})-(\ref{exp higgs action})
describes the SM field content plus a set of heavy gauge and fermionic excitations.
Its only exact gauge invariance is that of the SM gauge symmetry, as one
can easily check by noticing that the SM gauge fields couple to
the heavy fields only through SM covariant derivatives and field strengths. 
The (leading) terms of the SM Lagrangian, interpreted as the 
low-energy effective limit of the Lagrangian (\ref{entire exp action}) below 
$m_*$, $\tilde{m}_*$, $M_*$, then follow by setting to zero all terms 
in Eq.~(\ref{entire exp action}) that involve the heavy fields.
The SM Yukawa couplings, $Y$, are written in terms of the  
composite Yukawa couplings, $Y_{*}$, as follows:
\begin{equation}\label{SM Yukawa couplings}
\begin{split}
(Y_{u})_{ij} &= \sin \varphi_{q_{Li}} (Y_{*U})_{ij} \sin \varphi_{u_{Rj}} \\
(Y_{d})_{ij} &= \sin \varphi_{q_{Li}} (Y_{*D})_{ij} \sin \varphi_{d_{Rj}}\, ,
\end{split}  \qquad\quad
\begin{split}
(Y_{\nu})_{ij} &= \sin \varphi_{\ell_{Li}} (Y_{*N})_{ij} \sin \varphi_{\nu_{Rj}} \\
(Y_{e})_{ij} &= \sin \varphi_{\ell_{Li}} (Y_{*E})_{ij} \sin \varphi_{e_{Rj}}\, ,
\end{split}
\end{equation}
where there is no sum on repeated indices. Note that the SM Yukawas are off-diagonal and 
{\it hierarchical}, while those before the mass diagonalization are off-diagonal and 
{\it non-hierarchical}.
Using Eq.~(\ref{SM Yukawa couplings}), the mass-diagonalized Lagrangian 
(\ref{entire exp action}) can be written in terms 
of all the SM parameters (gauge and Yukawa couplings plus the Higgs quartic coupling
and mass term), as well as mixing angles for each SM particle and 
heavy-physics mass scales, $m_*, \tilde{m}_*, M_*^2$.

\subsection{Parameter Space Beyond the Standard Model}

Now that we have presented the diagonalized Lagrangian containing the SM and a set of heavy fields 
with which the SM particles interact, we comment briefly on the extended parameter space of the model.  
We began with a model described by the theoretical parameter space of the elementary and composite sectors, 
$\{g_{el}, g_*, Y_*, m, \tilde{m}, \Delta, \tilde{\Delta}, M_*, \lambda_H, \mu_H\}$.  After diagonalizing, 
we are left with a Lagrangian written in terms of the parameters of the Standard Model,
$\{g, Y, \lambda_H, \mu_H\}$, plus a set of mixing angles $\{\theta, \varphi, \tilde{\varphi}\}$ 
and masses of the new heavy states $\{ m_*, \tilde{m}_*, M_* \}$.
Thus, we can think of the mixing angles and the heavy masses
as new parameters which can be varied to explore the parameter space subject to the constraints:
\begin{equation}\label{constraints1}
g =  g_* \sin\theta \, , 
\end{equation}
\begin{equation}\label{constraints2}
(Y_{SM})_{ij}  =  \sin\varphi_{\psi_{L i}} (Y_*)_{ij}\sin\varphi_{\psi_{R j}} \, ,
\end{equation}
and our assumption that the composite sector is more strongly coupled than the elementary sector:
\begin{equation}\label{assumption}
g_*, Y_* \sim 1 - 4 \, .
\end{equation}

\section{EWSB}
\label{EWSB}

After EWSB, mass terms proportional to the Higgs vev
mix different states, and a further diagonalization is needed. In the electroweak 
unitary gauge, the charged and neutral gauge mass matrices are:  
\newpage
\vspace*{-1cm}
\begin{equation}
{\small
\begin{split}
 & \hspace{1.1cm} W^+_\mu \hspace{1.4cm}   W^{*\, +}_\mu \hspace{1.8cm} \tilde W^+_\mu  \\[0.2cm]
    M^2_{\pm} = \frac{v^2}{4} &
   \begin{pmatrix} 
 \gl_2^2  & \gl_2^2 \,\ds\frac{c_2}{s_2}  &  -\gl_1 \gl_2 \,\ds\frac{1}{s_1} \\[0.35cm]
 \gl_2^2 \,\ds\frac{c_2}{s_2}  & \ds\frac{4 M_{*2}^2}{c_2^2 v^2} +\gl_2^2 \,\ds\frac{c_2^2}{s_2^2}   
   & \ds -\gl_1 \gl_2\, \frac{c_2}{s_1 s_2} \\[0.35cm]
   -\gl_1 \gl_2 \,\ds\frac{1}{s_1}  &  \ds -\gl_1 \gl_2 \, \frac{c_2}{s_1 s_2}
   & \ds \frac{4 M_{*1}^2}{v^2} + \gl_2^2 \,\ds\frac{1}{s_2^2}
       \end{pmatrix}
       \begin{array}{l}
        W^-_\mu \\[0.5cm]  W^{*\, -}_\mu \\[0.5cm] \tilde W^-_\mu
       \end{array}
\end{split} 
}
\end{equation}
%
\vspace{0.4cm}
\begin{equation} \label{neutralM}
{\small
\begin{split}
 & \hspace{1.1cm} W^{3}_\mu \hspace{1.8cm}   {\cal B}_\mu \hspace{2.1cm}  W^{*\, 3}_\mu  
   \hspace{2.4cm} {\cal B}^{*}_{\mu} \hspace{2.6cm} \tilde {\cal B}_\mu \\[0.2cm]
    M^2_{0} = \frac{v^2}{4} &
   \begin{pmatrix} 
 \gl_2^2     & -g_Y\gl_2  & \gl_2^2\,\ds\frac{c_2}{s_2} & -g_Y\gl_2 \,\ds\frac{c_1}{s_1} 
                       &  -g_Y\gl_2 \,\ds\frac{1}{s_1}\sqrt{\frac{2}{3}}  \\[0.5cm]
 -g_Y\gl_2 & g_Y^2 & -g_Y\gl_2 \,\ds\frac{c_2}{s_2} & g_Y^2 \,\ds\frac{c_1}{s_1} 
                       & g_Y^2 \,\ds\frac{1}{s_1}\sqrt{\frac{2}{3}}  \\[0.5cm]
 \gl_2^2 \,\ds\frac{c_2}{s_2} & -g_Y\gl_2 \,\ds\frac{c_2}{s_2}  & \ds\frac{4 M_{*2}^2}{c_2^2 v^2} 
  + \gl_2^2 \,\ds\frac{c_2^2}{s_2^2} & -g_Y\gl_2  \,\ds\frac{c_1 c_2}{s_1 s_2} 
            & -g_Y\gl_2 \,\ds\frac{c_2}{s_1 s_2}\sqrt{\frac{2}{3}}  \\[0.5cm]
 -g_Y \gl_2\,\ds\frac{c_1}{s_1} & g_Y^2 \,\ds\frac{c_1}{s_1} & -g_Y\gl_2 \,\ds\frac{c_1 c_2}{s_1 s_2}
            & \ds\frac{4 M_{*1}^2}{c_1^2 v^2} + g_Y^2 \,\ds\frac{c_1^2}{s_1^2}
            & g_Y^2 \,\ds\frac{c_1}{s_1^2}\sqrt{\frac{2}{3}}  \\[0.5cm]
  -g_Y\gl_2 \,\ds\frac{1}{s_1}\sqrt{\frac{2}{3}}  &  g_Y^2 \,\ds\frac{1}{s_1}\sqrt{\frac{2}{3}} 
            & -g_Y\gl_2 \,\ds\frac{c_2}{s_1 s_2} \sqrt{\frac{2}{3}} 
            & g_Y^2  \,\ds\frac{c_1}{s_1^2}\sqrt{\frac{2}{3}}  &  \ds \frac{4 M_{*1}^2}{v^2} +  g_Y^2 \,\ds\frac{1}{s_1^2}\sqrt{\frac{2}{3}} 
    \end{pmatrix} 
       \begin{array}{l}
         W^{3}_\mu \\[0.7cm]   {\cal B}_\mu \\[0.7cm]  W^{*\, 3}_\mu  \\[0.7cm] 
         {\cal B}^{*}_{\mu} \\[0.7cm] \tilde {\cal B}_\mu
       \end{array}
\end{split} 
}
\end{equation}
where $v=246$~GeV, $s_{1,2}\equiv\sin\theta_{1,2}$, $c_{1,2}\equiv\cos\theta_{1,2}$,
and we have expressed Eq.~(\ref{neutralM}) in terms of $g_Y = \sqrt{3/5}\, g_1$ instead of $g_1$
to have a more easy-to-read expression.
The fermionic mass matrices for up and down states are:
\begin{align}
\begin{split}
 & \hspace{0.9cm} u_L \hspace{1.9cm} U_L \hspace{2.1cm} \tilde U_L \\[0.2cm]
    M_{U} =  &
   \begin{pmatrix}
     Y_u\, \ds\frac{v}{\sqrt{2}} & \ds\frac{c}{s}\, Y_u\, \ds\frac{v}{\sqrt{2}}& 0 \\[0.35cm]
     0 & m^Q_* & \ds s^{-1} Y_u\, \tilde s_u^{-1}\, \ds\frac{v}{\sqrt{2}} \\[0.35cm]
     Y_u\, \ds\frac{\tilde c_u}{\tilde s_u}\, \ds\frac{v}{\sqrt{2}} 
       & \ds\frac{c}{s}\, Y_u\, \ds\frac{\tilde c_u}{\tilde s_u}\, \ds\frac{v}{\sqrt{2}} & \tilde m_{*}^U
   \end{pmatrix}
   \begin{array}{l}
        u_R \\[0.5cm]  U_R \\[0.5cm] \tilde U_R
   \end{array}
\end{split}
\\[0.5cm]
%
\label{MD}
\begin{split}
 & \hspace{0.9cm} d_L \hspace{1.9cm} D_L \hspace{2cm} \tilde D_L  \\[0.2cm]
    M_{D} =  &
   \begin{pmatrix}
     Y_d\, \ds\frac{v}{\sqrt{2}} & \ds\frac{c}{s}\, Y_d\,  \ds\frac{v}{\sqrt{2}} & 0 \\[0.35cm]
     0 & m^Q_* & \ds s^{-1} Y_d\,  \tilde s_d^{-1}\, \ds\frac{v}{\sqrt{2}} \\[0.35cm]
     Y_d\, \ds\frac{\tilde c_d}{\tilde s_d}\, \ds\frac{v}{\sqrt{2}}
       & \ds\frac{c}{s}\, Y_d\, \ds\frac{\tilde c_d}{\tilde s_d}\, \ds\frac{v}{\sqrt{2}}  & \tilde m_{*}^D
   \end{pmatrix}
   \begin{array}{l}
        d_R \\[0.5cm]  D_R \\[0.5cm] \tilde D_R
   \end{array}
\end{split} 
\end{align}
where we have defined $s\equiv\sin\varphi_{u_L}$, $c\equiv\cos\varphi_{u_L}$; 
$\tilde s_{u, d}\equiv\sin\varphi_{u_R,d_R}$, $\tilde c_{u, d}\equiv\cos\varphi_{u_R,d_R}$, 
and $Y_u$, $Y_d$ are the up- and down-type SM Yukawa couplings
defined by Eq.~(\ref{SM Yukawa couplings}).

In general it is best to proceed numerically for any choice of model 
parameters in diagonalizing the above mass matrices, and rewriting 
the Lagrangian of Eq. (\ref{entire exp action}) in terms of the resultant mass-eigenstate 
fields, thereby obtaining the final interaction vertices for the 
bottom-line mass eigenstates. 
In the case of the gauge mass matrices, however, a reasonable approximation consists
in treating the EWSB corrections as small perturbations, and work at leading order.
This is consistent since the EWSB terms are smaller than  the
mass splitting in the heavy-heavy sector.
In the fermionic mass matrices this is not true in general, and a full diagonalization
is therefore needed. A further complication comes from the fact that each of the elements
of $M_U$, $M_D$ is actually a $3\times 3$ matrix in flavor space.
Depending on the flavor structure of the composite Yukawa matrices $Y_*$, 
which in turn must combine with the $\sin\varphi$ to yield realistic SM Yukawa 
couplings,
a full diagonalization might be needed only generation by generation, or 
worst case for the whole
$9\times 9$ fermionic matrices. The first possibility occurs if $Y_*$ is approximately
diagonal in flavor space, maybe as the result of a flavor symmetry of the composite sector.
No simplification is instead possible in the opposite, extreme case of anarchic $Y_*$,
that is if all entries of $Y_*$ are of the same order and large.
Quite interestingly, these two different realizations of flavor 
could be experimentally distinguishable at future colliders.
We will discuss this in Section~\ref{pheno}.

\section{Full $t_R$ Compositeness }
\label{full tR compositeness}

We have already said that the size of the SM Yukawa couplings is controlled, in our model,
by the size of the composite Yukawas $Y_*$ and by the degree of compositeness of the fermion mass 
eigenstates, Eq.~(\ref{SM Yukawa couplings}). In the particular case of the top quark, we have:
\begin{equation} 
\label{top}
Y_{top} =  \sin \varphi_{t_L} \left(Y_{*U33}\right) \sin \varphi_{t_R}\, .
\end{equation}
Simple inspection of this formula shows that the (large) SM top Yukawa coupling can be reproduced
only if the mixing angles of $t_R$ and $t_L$ are not both too small.
We are assuming that $(Y_{*U33})$ is not too strong, say 
$Y_{*U33} \lesssim 3$, both to stay in theoretical control of the composite 
sector and also because the electroweak $T$ parameter scales as $(Y_{*U33})^4$.
Now the mixing angles control the partial compositeness phenomenology of SM particles, 
yielding deviations from SM predictions. While present top quark tests 
pose no conflict with sizeable mixing angles, 
note that the electroweak symmetry implies 
\begin{equation} 
\varphi_{b_L} = \varphi_{t_L}, 
\end{equation}
and the couplings of $b_L$ are very well tested, in particular its coupling
to the $Z$, $g_{Lb}$.
Large values of the mixing angle $\varphi_{b_L}$ imply a sizable correction $\delta g_{Lb}$
to the $Zb\bar b$ vertex, as the calculation of Appendix~\ref{sec:ztobbbar} explicitly shows.

\begin{figure}[t]
\hspace{1cm}
\epsfig{figure=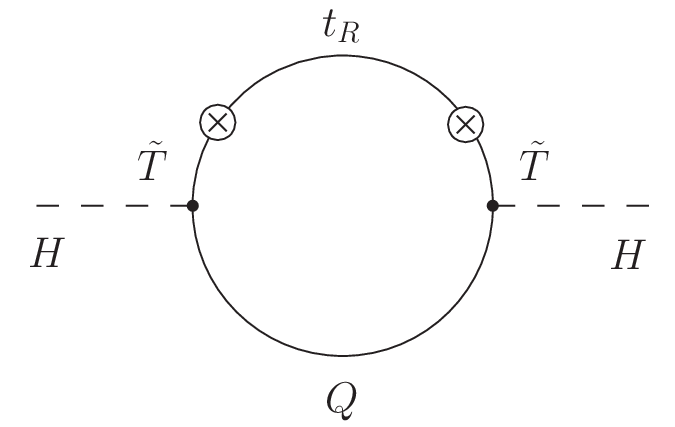,scale=0.6}
\begin{minipage}[c]{0.4\linewidth}
\vspace*{-4.2cm} \hspace{0.3cm}
$\ds \sim - \frac{3}{4 \pi^2} \left(Y_{*U33}\right)^{2} (\tilde m_*^T)^2 \sin^2\varphi_{t_R}
 \log\left(\frac{\Lambda^2}{\tilde m^{T\, 2}_{*}}\right)$
\end{minipage}
\caption{\it One-loop log divergent contribution to the Higgs mass squared 
from the virtual exchange of an elementary $t_R$. A circled cross denotes a $\Delta_{t_R}$ 
mass mixing. We used Eqs.~(\ref{R fermion mass diag}), (\ref{exp mass})
to set $\Delta_{t_R} = \tilde m_*^T \sin\varphi_{t_R}$, and the fact that additional 
$\Delta_{t_R}$ insertions on the $t_R$ elementary propagator are suppressed 
for divergent virtual momenta. }
\label{top loop in A}
\end{figure}

The best one can imagine in easing the tension 
between the constraint on $\delta g_{Lb}$ and Eq.~(\ref{top}) is to satisfy  
the latter with the smallest possible $\varphi_{t_L}$ and the largest possible 
$\varphi_{t_R}$.
This has an immediate cost, in that loops involving the elementary 
$t_R$, such as Fig.~\ref{top loop in A}, strongly correct the Higgs mass squared, with 
composite-strength couplings and without mixing-angle 
suppression. For $m_* >$ TeV, this implies a Higgs fine tuning at the 
$10$ percent level or worse. This problem reflects a breakdown in our 
philosophy. For $\sin\varphi_{t_R} \sim {\cal O}(1)$, the separation between 
an elementary sector weakly coupled to 
(and weakly mixed with) a stronger composite sector is lost.

These problems suggest a different scenario.
Rather than thinking of 
$t_R$ as an elementary field strongly coupled to the composites,
it makes better sense to posit that the $t_R$ is itself a full 
chiral member of the composite sector (with no elementary admixture), 
and its participation in the composite 
dynamics is an integral part of generating a light Higgs multiplet.~\footnote{
This now means that SM gauge anomalies cancel non-trivially between the 
new elementary and new composite sectors.} 
From this viewpoint, quantum loops involving only the
$t_R$ and other composites, Fig.~\ref{top loop in A}, do 
not contribute to our measure of fine tuning, as discussed in subsection 4.2.
With the $t_R$ now a full composite, the remaining elementary particles have 
reasonably small mixing angles suppressing radiative corrections to the Higgs 
mass squared, resulting in mild tuning. 

This is the physical picture we will assume from now on. The question is 
how to amend our present construction to take this into account. 
We do this by continuing to use Eq. (\ref{entire exp action}) 
to pursue the phenomenological implications of the model, 
but we simply take a composite-$t_R$ limit,
\begin{equation}
\label{limit}
\sin \varphi_{t_R} = 1 \, .
\end{equation}
As discussed above, Higgs fine tuning is then only measured with respect 
to sensitivity to mixing angles of the remaining elementary fields.

While Nature might have even heavier composites with the quantum numbers of 
the light Higgs, they are not required in our minimal model, and 
we have not included them. Similarly, now that  $t_R$ is a complete 
composite, which we allow to interact directly with the Higgs {\it before}
the diagonalization of Section~\ref{partial compositeness}, 
we will no longer need to retain the heavy Dirac composite 
excitation with the same quantum numbers, $\tilde{T}$, in the minimal model 
(although again Nature may possess such a state). A minor extra payoff of 
this is that the dominant contributions to fine tuning in the Higgs mass 
will turn out to be UV-finite, that is there are not even 
logarithmic divergences. Again, we 
can continue to utilize Eq.~(\ref{entire exp action}) for phenomenological investigation, 
but with the second limit in which we throw away terms containing $\tilde{T}$:
\begin{equation}
\tilde{T} \rightarrow 0 \, .
\end{equation}
In performing low-momentum calculations, we find it useful to work with the lagrangian before 
elementary/composite diagonalization, Eqs.~(\ref{ele action}), (\ref{comp action}) and (\ref{soft mix}). 
The equivalent   $t_R$-compositeness limit in this language
is given by
\begin{equation}
\tilde{T}_L\rightarrow 0\, , \qquad \tilde{T}_R\rightarrow t_R \, .
\end{equation}

\section{Higgs Fine Tuning and Finiteness}
\label{higgs fine-tuning and finiteness}

Let us discuss more in detail about the Higgs fine tuning in the minimal model 
with composite~$t_R$. As anticipated in the overview, we choose as a measure of tuning
the ratio of the desired Higgs mass squared, $m_h^2$, to the largest 1-loop correction 
sensitive to the mixing with the elementary sector,
\begin{equation} \label{ft}
\text{Fine tuning} \sim \frac{m_h^2}{\delta m^2_{h|mixing}}\, .
\end{equation}
Note that $m_h$ denotes the physical Higgs boson 
mass, not the mass parameter of the Higgs doublet.
The 1-loop correction to the Higgs mass squared from the composite sector alone 
is assumed to be reasonably well subsumed into the parameters of the 
tree-level potential, $V(H)$, and by assumption does not contribute to fine tuning, 
as discussed in subsection 4.2.

\begin{figure}[t]
\begin{center}
\epsfig{figure=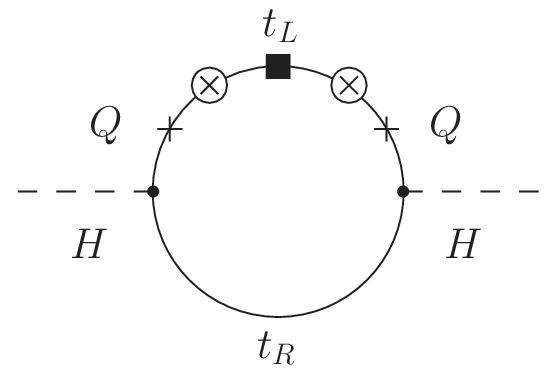,scale=0.6} \\
\epsfig{figure=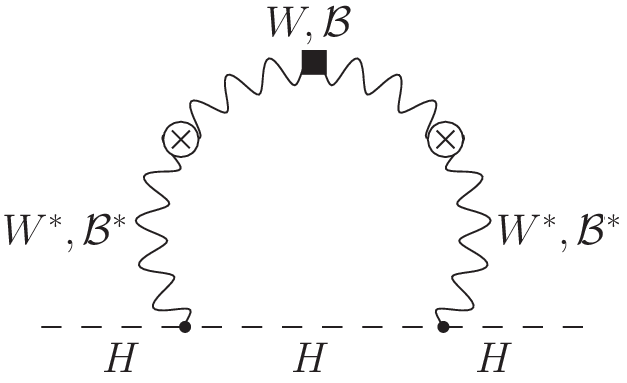,scale=0.6} \hspace{1cm}
\epsfig{figure=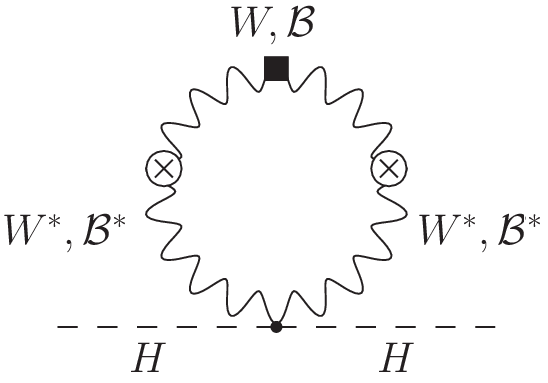,scale=0.6}
\caption{\it 
One-loop corrections to the Higgs mass squared
in the minimal model with composite $t_R$
adopting the elementary/composite basis. Black squares represent 
resummation of all higher elementary/composite mass-mixing.}
\label{higgs ft insertion}
\end{center}
\end{figure}

The corrections $\delta m^2_{h|mixing}$ from gauge and fermionic fields 
are most easily computed in the elementary/composite basis. The relevant Feynman diagrams are those
of Fig.~\ref{higgs ft insertion}. 
This choice of basis also makes manifest the central mechanism for cutting off 
the large quadratic divergences: the elementary fields only couple to the Higgs via their mass mixing
with the composite fields, and this implies enough propagators to make the loop integral convergent. 
Technically, this good behavior reflects the collective-breaking
structure of our model. Physically, it captures the UV form-factor 
suppression due to partial compositeness. 

The computation of the diagrams of Fig.~\ref{higgs ft insertion} is straightforward, with the only 
subtlety of resumming all possible elementary/composite mass insertions on 
the elementary propagator. In the case of the elementary $t_L$ field,
the effect of this resummation is that of modifying the propagator to:
\begin{equation}
\frac{i}{\pslash} \quad\longrightarrow\quad \frac{i}{\pslash} \cdot
 \frac{1}{ 1 - \ds\frac{\Delta_{t_L}^2}{p^2 - \left(m^{Q}\right)^2}  } \, .
\end{equation}
The resummed elementary gauge propagator is instead
\begin{equation}
\frac{-i}{p^2} \cdot \frac{1}{1 - \ds\frac{M_*^2 \tan^2\theta }{p^2 - M_*^2} } 
 \left( \eta_{\mu\nu} - \frac{p_\mu p_\nu}{p^2}  \right)\, ,
\end{equation}
while the (un-resummed) $W^*$, $B^*$ propagator to be used to compute the gauge diagrams
of Fig.~\ref{higgs ft insertion} is the usual one for a vector of mass $M_{*1}$, $M_{*2}$.
We thus obtain, respectively for the gauge and fermion contribution to the physical 
Higgs mass squared (after rotating to the Euclidean),
\begin{align}
\label{gauge loop}
&\delta m^2_{h|gauge} =
- \frac{9\, \gl^2_2}{16\pi^2}  \int^\infty_0 \!\!\! dp\; p\cdot  {\cal F}_{gauge}(p^2, \theta_2, M_{*2}) 
   + \frac{1}{3}\cdot \frac{3}{5} \left (\theta_2, \gl_2, M_{*2} \leftrightarrow \theta_1,\gl_1, M_{*1} \right )\, , 
\\[0.2cm]
&{\cal F}_{gauge}(p^2, \theta_2, M_*) =
  \frac{M^4_*/{\cos^2\theta_2}}{\left ( p^2 + M^2_* \right ) \left( p^2 + M^2_*/\cos^2\theta_2 \right) }\, ,  \\[0.7cm]
&\delta m^2_{h|fermion} =
   \frac{3}{2\pi^2}\, Y_{top}^2 \int^\infty_0 \!\!\! dp\; p\cdot
  {\cal F}_{fermion}(p^2,\varphi_{t_L},m_{*33}^Q)\, ,  \\[0.2cm]
&{\cal F}_{fermion}(p^2,\varphi_{t_L},m_{*33}^Q) =
  \frac{(m_{*33}^{Q})^4 \cos^2\varphi_{t_L}}{\big( p^2 + (m_{*33}^{Q})^2 \cos^2\varphi_{t_L} \big)
  \big( p^2 + (m_{*33}^{Q})^2 \big) }\, . 
\end{align}
All formulas have been expressed in terms of physical mass parameters and mixing angles, and we
have neglected the tree-level Higgs mass for simplicity.
The factor of $3/5$ in Eq.~(\ref{gauge loop}) is due to our normalization of the hypercharge.
The form factors ${\cal F}_{fermion}$, ${\cal F}_{gauge}$ 
cut off the quadratic divergences for
virtual momenta much larger than the mass of the heavy excitations; setting them to 1 gives 
the divergent SM results.
By approximating $\cos\varphi_{t_L} = 1$, $\cos\theta_{1,2} = 1$ in the form factors, 
one obtains
\begin{equation} \label{dmH}
\delta m^2_{h|gauge} \simeq -\frac{3}{32\pi^2} 
 \left (3\, \gl^2_2 \, M^2_{*2} + \gl_Y^2 M^2_{*1} \right ) \, ,
 \qquad
\delta m^2_{h|fermion} \simeq  \frac{3}{2\pi^2}\, \frac{m_t^2}{v^2} \, (m_{*33}^Q)^2 \, .
\end{equation}

\section{Viable Parameter Space}
\label{viable parameter space}

While the leading terms in the SM EFT below $m_*, M_*$ are given by 
simply setting to zero all heavy fields in Eq. (\ref{entire exp action}), the
body of precision electroweak and flavor data also constrains higher-dimension 
operators obtained by more carefully integrating out the heavy physics. 
The most important constraints come from oblique corrections encoded in the
Peskin-Takeuchi $S$ and $T$ parameters~\cite{PT}, modifications of the $Zb_L \bar b_L$ coupling, 
$\delta g_{Lb}$, and flavor-changing four-fermion operators.  
The leading contributions to $S$, $T$ and $\delta g_{Lb}$ 
due to the exchange of the heavy resonances are computed in the Appendix,
Eqs.~(\ref{S par}), (\ref{T par loop}) and (\ref{Zbb}).
These simple formulas can be used to identify 
the viable parameter space, determining thereby where and how
new physics can be produced explicitly in collider searches. 
Flavor physics experiments constrain the 
parameter space in more complex ways, sensitive to $Y_{*U,D}$, and 
can be neglected in a simple, model-independent analysis.
Broadly speaking, there is a generalized GIM mechanism saving the model from catastrophic 
failure in regard to flavor-changing rare processes, which is illustrated in Appendix~\ref{FCNC}.
Various dedicated analyses have been carried out
in the context of full higher-dimensional 
warped models~\cite{flavor}. A rough conclusion appears to be that 
$M_*$ should be above roughly $2$ TeV, in particular $M_{*3}$, in order for 
reasonable flavor ansatze to survive. This agrees and is consistent with the 
constraints imposed by the electroweak observables, 
which we now want to discuss in some detail.\footnote{
Precision electroweak effects are but a virtual shadow of new physics, 
which  can easily be sensitive to modest non-minimal physics 
which we cannot anticipate in advance. Therefore in 
collider searches for new physics, precision data should be consulted to 
give a ballpark guide to the viable parameter space, rather than trusting the 
sharp boundaries that emerge in analysing any particular model.
For example, the constraint from $Zb\bar b$ can be strongly relaxed 
if the $SU(2)_L\times SU(2)_R$ symmetry of the composite sector is enlarged
to include a discrete LR parity, and additional composite states are 
added~\cite{Agashe:2006crp}.}

\begin{figure}[t]
\begin{center}
\epsfig{figure=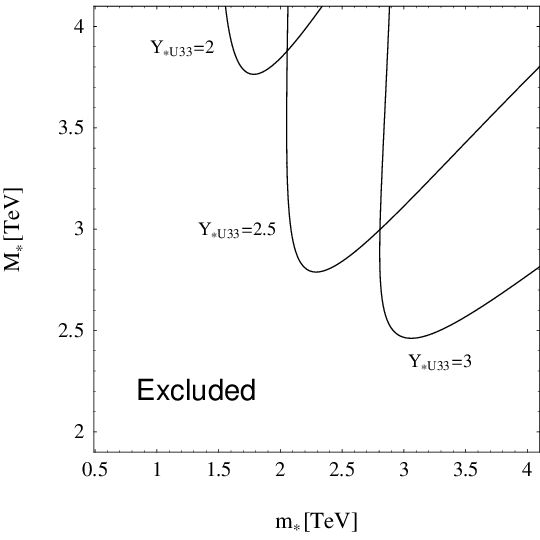, scale=0.8} \hspace{0.8cm}
\epsfig{figure=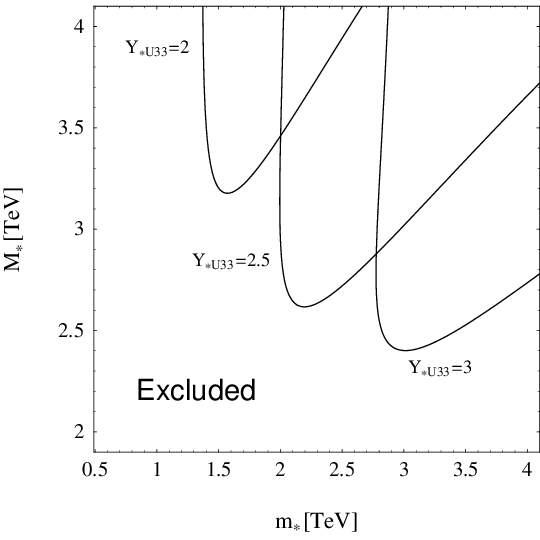, scale=0.8}
\caption{\it Exclusion curves in the plane $(m_*,M_*)$ for $Y_{*U33}=2,2.5,3$ and 
$Y_{*D33}/Y_{*U33}=1/2$ (left plot) or $Y_{*D33}/Y_{*U33}=1/3$ (right plot). 
The area below each curve is excluded at $99\%$ CL by
a combined fit to the electroweak observables $S$, $T$ and~$\delta g_{Lb}$.
}
\label{EWcp}
\end{center}
\end{figure}

Let us make the simplifying hypothesis of universal masses $M_*$, $m_*$ and 
composite couplings $g_*$. Equations (\ref{S par}), (\ref{T par loop}) and (\ref{Zbb})
then express the three precision observables $S$, $T$ and $\delta g_{Lb}$ as
functions of $m_*$, $M_*$, $Y_{*U33}$, $Y_{*D33}$, and $g_{*}$.
\footnote{We use the formula of the top Yukawa coupling, Eq.~(\ref{top}),
to express $\sin\varphi_{t_L}$ in terms of $Y_{*U33}$, and hence
$m^Q_{33}$ in terms of $m^Q_{*33}$ and $Y_{*U33}$.}
For definiteness, we set $g_* = 3$ and vary the remaining parameters.
The body of precision data (see for example~\cite{LEPEWWG,Barbieri:2004qk}) 
can then be used to constrain the plane $(m_*,M_*)$ for fixed $Y_{*U33}$ and $Y_{*D33}$,
provided the Higgs mass $m_h$ is also specified.
The results are shown in Fig.~\ref{EWcp}, where we used $m_h = 250$ GeV.
\footnote{In detail, we perform a $\chi^2$ test using the
fit to $S$, $T$ and $\delta g_{Lb}$ of Ref.~\cite{Agashe:2006c}
with $m_h = 250$ GeV, and impose a $99\%$ CL bound.}

One can see that there is a sizable portion of parameter space that is
not ruled out and that should be accessible at the LHC, as we will discuss
more in detail in the next section.
The constraint is stronger on $M_*$, mainly due to the $S$ parameter, while
the fermion mass $m_*$ can be lighter. Lighter masses for the fermionic resonances
are also preferred to keep the residual fine tuning in the Higgs mass small.
The estimate of Eq.~(\ref{ft}), together with Eq.~(\ref{dmH}), shows that the level of 
cancellation in the one-loop correction to the Higgs mass squared, required to obtain
$m_h = 250$ GeV, is not worse than $\sim 10\%$ for $m_*$ up to 3~TeV.
This means that the viable region of Fig.~\ref{EWcp}
is also a natural place for the parameters of our model.

\section{Phenomenology Highlights and Challenges}
\label{pheno}

The minimal scenario of partial compositeness 
that emerges from the electroweak precision tests
is particularly challenging for the LHC.
Vector excitations are to be at 2-3 TeV or heavier,
while fermions can be lighter, as
also suggested by fine-tuning considerations.
The new states are expected to couple strongly to the third-generation 
SM quarks, but weakly to the light fermions.
This makes them difficult to discover at a hadron collider,
though there are specific processes that are particularly promising.

Partial compositeness predicts a quite well determined 
pattern of new physics at the LHC, whose key features 
are robust and do not depend on the details of the models.
Perhaps the most relevant exception to this rule comes in the
phenomenology of the heavy fermions, where two qualitatively
different scenarios can arise, depending on whether flavor mixing 
effects in the composite sector are large or not.
We will start by considering first the case in which flavor-changing
effects in the composite Yukawa matrices $Y_*$ are small, and
then see how the phenomenology changes by relaxing this hypothesis.

\subsection{Heavy gauge boson production and decays}

The most effective strategy to discover the new particles can be deduced
by looking at the magnitude of their couplings in the Lagrangian (\ref{entire exp action}).
Heavy excitations of the SM 
gauge bosons, $\rho^*$, couple to the SM fermions with strength
\begin{equation*}
\gl \left( \sin^2\varphi \cot\theta - \cos^2\varphi \tan\theta \right)\, ,
\end{equation*}
as one can see from the second line of Eq.~(\ref{exp fermion action}).
In the case of light SM fermions
the first term is highly suppressed, since $\varphi\ll 1$, and the whole
interaction is accounted for by the second term. The latter 
is due to the (universal) $A-\rho$ mixing and it is still suppressed
by a factor $\tan\theta \simeq (\gl/g_*)$ compared to SM couplings. 
By contrast, third-generation quarks
have a large mixing angle $\varphi$ and the first term dominates.
The coupling of $\rho^*$ to the longitudinal polarizations of the SM weak bosons 
is also strong, the latter being composite degrees of freedom. 
It can be extracted, in the unphysical basis
and using the Equivalence theorem, from the second line of
Eq.~(\ref{exp higgs action}): $g\cot\theta \simeq g_*$.
Producing and detecting the heavy vectors will be challenging for the LHC
due to their small couplings to light fermions.
Single production mainly proceeds via Drell-Yan scattering
\begin{equation*}
q\, \bar q \longrightarrow \rho^* \, .
\end{equation*}
Electroweak heavy excitations $W^{*\, \pm}$, $W^{*\, 3}$, $\cal{B}^* $~\footnote{Here and in the following 
we classify the new vectors using their $SU(2)_L$ quantum numbers, since the EWSB
effects are small, see section \ref{EWSB}.}
can also be produced via weak boson fusion, as in technicolor theories.
Despite the large coupling to longitudinal $W$'s and $Z$'s,
this process is subdominant for large masses $M_*\gtrsim$ 2-3 TeV, as well
as other production mechanisms like gluon fusion, $t\bar t$ and $t\bar b$
associated productions.~\footnote{K.Agashe, private communication.} 
Once produced, the heavy vectors will mainly decay to pairs of third-generation SM quarks,
$t\bar t$, $b\bar b$ or $t\bar b$, and pairs of longitudinally polarized SM vector bosons,
$Z_L h$, $W_L^+ W_L^-$ or $W_L^\pm Z_L$.
Decays to light SM quarks and leptons will instead be rare.
When kinematically allowed, also decays to one SM top or bottom plus one excited top
or bottom quark ($T\bar t$, $T\bar b$ etc.), or even to any two heavy fermions will be important,
thanks to their large couplings to the excited vectors (first terms of third and fourth
lines in Eq.~(\ref{exp fermion action})).

In  the case of the electroweak heavy vectors  $W^{*\, \pm}$, $W^{*\, 3}$, $\cal{B}^* $,
it is very useful to compare our scenario with Little Higgs (LH) models, for which numerous and
detailed studies exist in the 
literature~\cite{Burdman:2002ns,Han:2003wu,Perelstein:2003wd,Azuelos:2004dm,Han:2005ru}.
In fact, the structure of the gauge sector of LH models  based on product groups
(and without T-parity) is similar to that prescribed by partial compositeness:
the product of two $SU(2)\times U(1)$ groups is broken down spontaneously 
to the diagonal subgroup (the latter being identified with the SM symmetry), and the SM
fermions transform under only the first gauge group.
This is analogous to the elementary/composite mixing of sec.~\ref{partial compositeness}, 
considering that the SM light fermions are almost completely elementary, and as such their couplings to the
composite gauge bosons are strongly suppressed.
In particular, the mixing angles between the two $SU(2)\times U(1)$ groups of the LH theories,  
$\theta_{LH}$, directly map into our parameters $\theta$.~\footnote{
More precisely, the correspondence is $\theta_{LH} = \theta$
in the convention of Ref.~\cite{Perelstein:2003wd}, while $\theta_{LH} = \pi/2 - \theta$  
in the notation of Refs.~\cite{Burdman:2002ns,Han:2003wu,Azuelos:2004dm,Han:2005ru}.
} 
This implies that in our model the production cross section for $W^{*\, \pm}$ and $W^{*\, 3}$ (as well as 
their decay widths to light fermions), will be the same as that for the heavy vectors of Little Higgs models
(see refs.~\cite{Burdman:2002ns,Han:2003wu}).
\footnote{
Notice also that the electroweak precision data from LEP
favor the region of parameter space of Little Higgs models
where the second $SU(2)\times U(1)$ becomes strong, $\tan\theta_{LH} \ll 1$,
and four-fermion contact interactions are thus suppressed~\cite{Barbieri:2004qk,LHprec} (for more references on LH models and precision tests, see the review ~\cite{Schmaltz:2005ts}).
In other words, the region of parameter space where LH models
pass the precision tests corresponds to that motivated by the 
partial compositeness paradigm.}
On the other hand, the pattern of decays will be quite distinct in the two cases.
In LH theories the couplings of $W^*$ to the various SM fermions
are  predicted to be either strictly universal (in product group models),
or comparable in size (in simple group models).
Partial compositeness, instead, predicts larger couplings 
to the third generation SM quarks, so that 
$tt$, $bb$ and $tb$ channels will have larger branching ratios.
For example, in the case of $W^{*\, 3}$ one has (neglecting the SM fermion masses and
the degree of compositeness of the SM leptons and light quarks):
\begin{equation} \label{Gammas}
\begin{split}
\Gamma(W^{*\, 3}\to  q \bar q) = 3\, \Gamma(W^{*\, 3}\to l \bar l)
 = & \, \frac{g_2^2 M_*}{32\pi} \tan^2\theta_2\, ,  \\[0.15cm]
\Gamma(W^{*\, 3}\to t \bar t) = \Gamma(W^{*\, 3}\to b \bar b) =& \, \frac{g_2^2 M_*}{32\pi}
 \left(\sin^2\varphi_{t_L} \cot\theta_2 - \cos^2\varphi_{t_L} \tan\theta_2 \right)^2 \, , \\[0.15cm]
\Gamma(W^{*\, 3}\to Z h) = \Gamma(W^{*\, 3}\to W^+ W^-)
 =& \, \frac{g_2^2 M_*}{192\pi} \cot^2\!\theta_2\, , \\[0.15cm]
\Gamma(W^{*\, 3}\to T \bar t) = \Gamma(W^{*\, 3}\to B \bar b) 
 =& \, \frac{g_2^2 M_*}{32\pi }\, \frac{\sin^2\varphi_{t_L} \cos^2\varphi_{t_L}}{\sin^2\theta_2 \cos^2\theta_2}  
        \left( 1- \frac{1}{2} \frac{m_*^2}{M_*^2} - \frac{1}{2} \frac{m_*^4}{M_*^4} \right)  \\
 &\times  \left(1- \frac{m_*^2}{M_*^2} \right)  \, , \\[0.15cm]
\Gamma(W^{*\, 3}\to \chi_q \bar \chi_q) 
 =& \, \frac{g_2^2 M_*}{32\pi} \, \Big\{ \!\left[ \left(\cos^2\!\varphi_{t_L}\cot\theta_2 
     - \sin^2\!\varphi_{t_L}\tan\theta_2 \right)^2 \! + \cot^2\!\theta_2 \right]  \\
   & \times \left( 1 - \frac{m_*^2}{M_*^2}  \right) + 6 \frac{m_*^2}{M_*^2}  \left(\cos^2\!\varphi_{t_L}\cot^2\!\theta_2 
     - \sin^2\!\varphi_{t_L} \right)\!\Big\} \\
   & \times \sqrt{1-4\frac{m_*^2}{M_*^2}} \\[0.15cm]
 \Gamma(W^{*\, 3}\to \chi_l \bar \chi_l) = & \, \frac{g_2^2 M_*}{96\pi} \cot^2\!\theta_2 \left(2+4\frac{m_*^2}{M_*^2} \right) 
       \sqrt{1-4\frac{m_*^2}{M_*^2}}     \, ,
\end{split}
\end{equation}
%
where $q$ ($l$) stands for any SM light quark (lepton) and $\chi_q = U_i, D_i$, $\chi_l = N_i, E_i$
for any flavor $i$.
The first identity in the third equation follows from the Equivalence theorem:
as pointed out above, final SM gauge bosons are almost completely longitudinally
polarized. 

Naively, the LHC discovery reach in the channels $l\bar l$, $l\nu$ and $Zh$, $WW$, $WZ$
will be similar, though not identical, to that derived in Ref.~\cite{Azuelos:2004dm} 
for the Littlest Higgs model~\cite{Arkani-Hamed:2002qy}. 
In particular, the larger couplings to pairs of SM bosons 
and third-generation SM quarks imply a smaller branching fraction to the clean leptonic 
channels $l\bar l$.
Despite the large branching ratio, $tt$ and $bb$ events will be
certainly challenging to isolate over the huge SM background. On the other hand, 
detecting a large violation of flavor universality, as predicted in our model,
would represent a first significant hint of partial compositeness.
The $Tt$, $Bb$, $Tb$, $Bt$ channels should be easier
to detect over the background, thanks to the richer final state.~\footnote{As we will 
discuss in detail below, heavy fermions mainly decay to a third-generation SM quark 
plus a longitudinally polarized gauge boson.}
If kinematically allowed, the decays to two heavy fermions  would be instead
extremely spectacular , especially
those of $W^{*\, \pm}$, $W^{*\, 3}$, $\cal{B}^* $ to a pair of heavy leptons, which would translate to 
quite distinctive leptonic final states.

Ideally, one would like to measure several of these channels, in order to extract all the 
parameters and test the model.
In the case of the $W^{*\, 3}$, for example, one can extract $\theta_2$ and $\varphi_{t_L}$ by counting the
number of events in any two of the final states of Eq.~(\ref{Gammas}); measuring a third channel 
then allows one to test the structure of the model.
This ideal strategy must however confront with the difficulty of the actual experimental
measurements. Quite likely, disentangling and measuring all the various parameters will be a hard task.
A minimal strategy to verify the model after the discovery and discriminate it
from other scenarios could be checking the relation:
\begin{equation} \label{pcsumrule}
 g_{HH\rho^*}^2 = \left( g\cot\theta\right)^2 = g_{HH\rho^*\rho^*} \, ,
\end{equation}
which is a direct consequence of the Higgs compositeness plus
partial compositeness in the gauge sector.
Similar tests have been also proposed for LH theories, see Refs.~\cite{Burdman:2002ns,Han:2003wu}.
The second equality in Eq.~(\ref{pcsumrule}) is certainly harder to verify,
due to the difficulty in measuring the coupling $g_{HH\rho^*\rho^*}$.
The latter can be extracted from the associated
production $q\bar q \to \rho^* \to \rho^* V$, with $V=h,W_L,Z_L$.

Another unique feature of the partial compositeness scenario, as compared to 
other models like LH theories, is the existence of heavy excitations of the gluon,
and the $\tilde \rho$ vectors.
The former will have a larger cross section than the neutral weak excitations
and will decay exclusively to $tt$, $bb$ and $Tt$, $Bb$ final states (and to two
excited quarks if kinematically allowed).
Very recently, Ref. \cite{k} has made an in-depth study of the $tt$ channel 
and shown how to exploit a left-right polarization asymmetry, deriving from  
$\varphi_{t_R} > \varphi_{t_L}$, to efficiently extract the signal above 
SM background for gluon excitations of several TeV.
 As explained above, we expect the $Tt$, $Bb$ channels
to also be favorable for detection, 
though a detailed study is needed.
The  heavy vectors $\tilde \rho$ are instead much more difficult to discover at the LHC
because in their case there is no analog of the $A-\rho$ mixing, and their couplings
to the light fermions will be extremely suppressed.
The associated production $q\bar q\to \rho^* \to \tilde\rho V$,
where $V = h, W_L, Z_L$, via a virtual or possibly real $\rho^*$, seems to be
the most promising process, though a detailed study is again required.

\subsection{Heavy fermions  production and decays}

If they are light enough, heavy excitations of the SM quarks will be pair produced 
at the LHC via QCD interactions:
\begin{equation}
 gg \, , \, q\bar q \to \chi\bar\chi\, ,
\end{equation}
and similarly for $\tilde\chi$.
The cross section of these processes is completely determined as a function of the 
mass $m_*$ of the new particle, and falls off quickly as $m_*$ increases
(see for example Refs.~\cite{Han:2003wu,Azuelos:2004dm}).
An additional contribution 
comes from the exchange of a gluon excitation (similar 
diagrams with the exchange of a $W^*$ or a $\cal{B}^* $ are also possible, but smaller):
\begin{equation}
q\bar q \to G^* \to \chi\bar\chi\, , \tilde\chi \bar{\tilde\chi}\, ,
\end{equation}
where  $G^*$ can be real, if kinematically allowed.
As discussed in the previous section, the electroweak precision tests
directly constrain the mass of the top and bottom quark excitations, 
disfavoring values smaller than roughly $1.5$ TeV.
For such heavy masses, pair production of these states becomes small at the LHC.
On the other hand, these same states couple strongly to their SM counterpart,
and can be singly produced.

In the case of $\tilde T$ and $\tilde B$, an important role is played by 
$b W$ and $b Z$ fusion~\cite{Willenbrock:1986cr}, 
where a bottom quark from one proton scatters off a longitudinal $W$ or $Z$ radiated
by a quark from the other proton:
\begin{equation} \label{single1}
\begin{split}
b_L W_L &\to \tilde T\, , 
 \qquad\quad \lambda_{\tilde T} = Y_{*U 33} \sin\varphi_{b_L} \cos\varphi_{t_R} =
 Y_{top} \cot\varphi_{t_R} \, , \\[0.1cm]
b_L Z_L &\to \tilde B\, , \qquad\quad \lambda_{\tilde B} = Y_{*D 33} \sin\varphi_{b_L} \cos\varphi_{b_R} =
  Y_{top}\, \frac{Y_{*D 33}}{Y_{*U 33}}\, \frac{\cos\varphi_{b_R}}{\sin\varphi_{t_R}} \, .
\end{split}
\end{equation}
The same production mechanism has been discussed and studied in the context
of Little Higgs models~\cite{Han:2003wu,Perelstein:2003wd,Azuelos:2004dm}. 
In order to make the comparison with the LH results 
easier, in Eq.~(\ref{single1}) we have written the couplings relevant to each process,
(they can be deduced from Eq.~(\ref{exp higgs action}) by making use of the Equivalence 
theorem).~\footnote{For comparison, the  heavy $SU(2)_L$ singlet predicted by the Littlest
Higgs model couples with strength $\lambda = Y_{top}\, \lambda_1/\lambda_2$, where
$\lambda_1/\lambda_2 \sim 1$.}
If $t_R$ is strongly coupled to the composite sector, $\sin\varphi_{t_R}\simeq 1$,
then $\lambda_{\tilde T}$ is small and the single production of $\tilde T$
is suppressed. In fact, in the extreme limit in which $t_R$ is part of the composite sector,
a heavy state $\tilde T$ is not required and we therefore 
removed it from our minimal 
model in Section 7.
The production rate of $\tilde B$ is instead expected to be sizable,
since its coupling is large for $t_R$ composite and $b_R$ almost elementary:
$\lambda_{\tilde B} \simeq Y_{top}\, (Y_{*D 33}/Y_{*U 33}) \approx 1$.
In the case of $T$ and $B$, single production via $b W$, $b Z$ fusion will be
extremely small if $b_R$ is almost elementary, and its couplings to the new states are
weak. 
The analogous processes initiated by a $t_R$,
\begin{equation} \label{single2}
t_R Z_{L} \to  T\, ,\qquad t_R W_{L} \to  B\, ,
 \qquad \lambda_{T} = \lambda_{B} = Y_{*U 33} \sin\varphi_{t_R} \cos\varphi_{t_L} 
 =  Y_{top} \cot\varphi_{t_L} \, ,
\end{equation}
might be large enough to be seen at the LHC, since
the small top quark content of the proton can 
be compensated by the large coupling~\cite{Agashe:2005cp}: 
$\cot\varphi_{t_L} \gg 1$ for $t_R$ composite.
A detailed analysis is however required.
Single production of $B$'s also proceeds 
via $g\, b_L\to B$, as the effect of the dimension-5 operator
$g_{\mu\nu} \bar B \sigma^{\mu\nu} b$  generated by loops of top quarks.
Finally, the associated production of  a heavy $\chi= T, B$ or $\tilde\chi = \tilde T, \tilde B$ 
together with a SM top or bottom quark is also possible ($\psi = t,b$):
\begin{equation}
q\bar q \to \rho^* \to \chi \psi\, , \tilde\chi \psi\, ,
\end{equation}
where $\rho^* = G^*, W^*, \cal{B}^* $ can be real.
It is worth stressing that these processes, as well as all the other 
single production mechanisms, are
strongly suppressed in the case of the excitations of SM light quarks, due to their
small couplings to the composite sector. For them, only pair production
will be viable.
The same conclusion also applies to heavy leptonic resonances, 
with the difference that they can only be pair produced via  the exchange
of $SU(2)_L\times U(1)_Y$ carriers.~\footnote{Clearly,  
there is no analog of the processes (\ref{single1}) and (\ref{single2}) in the case of the 
heavy leptons, though they could be singly produced in the decays 
of $W^*$ and $\cal{B}^* $ if they had large couplings.}

The decays of $T$, $B$, $\tilde T$, $\tilde B$  proceed through the same interaction vertices and couplings
responsible for their production. The $SU(2)_L$ singlets $\tilde T$ and $\tilde B$
decay to a top or bottom quark plus a longitudinal SM boson or a Higgs, with
branching ratios fixed by the Equivalence theorem:
\begin{equation} \label{GammasTtilBtil}
\begin{split}
\Gamma(\tilde T\to t h) = \Gamma(\tilde T\to t Z) = \frac{1}{2}\, \Gamma(\tilde T\to b W)
  & = \frac{\lambda_{\tilde T}^2}{32\pi}\, \tilde m_*\, , \\
\Gamma(\tilde B\to b h) = \Gamma(\tilde B\to b Z) = \frac{1}{2}\, \Gamma(\tilde B\to t W)
  & = \frac{\lambda_{\tilde B}^2}{32\pi}\, \tilde m_*\, .
\end{split}
\end{equation}
In addition to the above channels, 
the $\tilde T$ or $\tilde B$ can also decay to a $T$, $B$
if kinematically allowed.
If the difference is mass between the two resonances is not too small, 
the eventual phase-space suppression can be compensated by 
the larger coupling involved: $Y_{* U,D}$.
For example, if sufficiently heavier, a $\tilde B$ could decay to a $T$ ($B$) by emitting
a $W_L$ ($Z_L$). This would represent a source of $T$'s and $B$'s, which might
be otherwise difficult to produce, as we have seen. Furthermore, 
if $\tilde T$, $\tilde B$
undergo a decay chain instead of directly decaying to SM quarks, this leads to richer final states,
which will be presumably easier to isolate over the SM background.

As already noticed in the literature~\cite{delAguila:1989ba,Azuelos:2004dm,Han:2005ru}, 
detecting the neutral-current decays of Eq.~(\ref{GammasTtilBtil})
would distinguish $\tilde T$ and $\tilde B$ from a fourth generation of quarks, though 
it would not be a smoking gun of partial compositeness.
More peculiar to our scenario are the decays of $T$ and $B$: in the motivated case of
$b_R$ almost elementary, the only unsuppressed channels are:
\begin{equation} \label{GammasTB}
\Gamma(T\to t h) = \Gamma(T\to t Z) = \frac{\lambda_{T}^2}{32\pi}\,  m_*\, ,
\qquad \Gamma(B\to t W) =  \frac{\lambda_{B}^2}{16\pi}\,  m_*\, ,
\end{equation}
plus eventually the analogous decays to $\tilde T$, $\tilde B$, if kinematically allowed.
Detecting only final states with top and no bottoms quarks from $T$ and $B$
would then be compelling evidence of partial compositeness and its mechanism
for explaining the hierarchies in the SM Yukawas.
On the other hand, directly checking the mechanism responsible for the cancellation
of the quadratic divergence in the top loop will be challenging. A possible test in 
the case of $t_R$ composite consists in verifying the relation
\begin{equation}
\lambda_T = Y_{top} \cot\varphi_{t_L}\, ,
\end{equation}
by using the value of $\varphi_{t_L}$ extracted, for example, from the decays of $\rho^*$.
Determining $\lambda_T$, however, is not easy: 
a measurement from the total decay width is perhaps not possible, unless $\lambda_T$ is very large, 
due to the limited calorimeter resolution.
One could use the production rate in the $tW$, $tZ$ fusion channels, if they are observed at LHC,
although the theoretical error due to the uncertainties in the parton distribution
functions will be probably large.

The scenario we have described so far assumes that flavor mixings in the composite
sectors, induced by the off-diagonal entries of $Y_*$ in flavor space, are small.
If this is the case, another 
evidence for partial compositeness could come from the production of the 
excitations of light quarks and leptons. 
Under the hypothesis of small flavor mixing in the composite sector,
these heavy states are expected to be 
narrow resonances: they will 
decay to their SM counterpart plus a longitudinally weak boson or a Higgs, with
much smaller couplings compared to those of Eqs.~(\ref{GammasTtilBtil}), (\ref{GammasTB}).
Most likely, flavor-changing decays to top and bottoms will also have sizable
branching ratios, since the flavor suppression can be compensated by the larger 
coupling of the top and bottom to the composites.
In the leptonic sector, pair production of the heavy leptons
would give rise to spectacular signatures,
by decaying to ultra-energetic SM leptons with essentially no SM background.

In the opposite, extreme limit of anarchic $Y_*$, flavor mixing effects among the
composite fermions will be large. 
Single production of the heavy quark
excitations will still proceed as discussed above, namely through a bottom or possibly
a top quark from the proton, but the heavy $T$, $B$, $\tilde T$, $\tilde B$
produced in this way will easily mix with the excitations of the light quarks.
These latter, on the other hand, will not decay anymore to light SM quarks, since
it will be more convenient for them to change their flavor and decay to tops and bottoms.
This scenario will then be distinguishable from the previous one for the absence
of final states with light jets arising from the pair production of the first and second 
family quark excitations. 
Similar considerations in the case of pair production of heavy leptons show that these
will mainly decay to taus, while muons and especially electrons in the final state will be rare.

We conclude by mentioning that the quark heavy excitations
will produce also indirect effects, some of which
might be spectacular, depending on the mass of the new particles.
Important examples are a possible large modification to the Higgs production
cross section via gluon fusion, and large shifts in the couplings
of the top quark to the SM weak bosons.

\section*{Acknowledgments}
We would like to thank Kaustubh Agashe, David E. Kaplan, Markus Luty, 
Alex Pomarol, Matthew Schwartz and Jay Wacker for useful discussions.  
We thank Stefania De Curtis for pointing out an error in the first version of the paper.
This work was supported in part by the  National Science Foundation grant NSF-PHY-0401513 
and in part by  the Johns Hopkins Theoretical Interdisciplinary Physics and Astrophysics Center.

\appendix

\section{Generalized GIM mechanism for FCNCs}
\label{FCNC}

\begin{figure}[t]
\begin{center}
\epsfig{figure=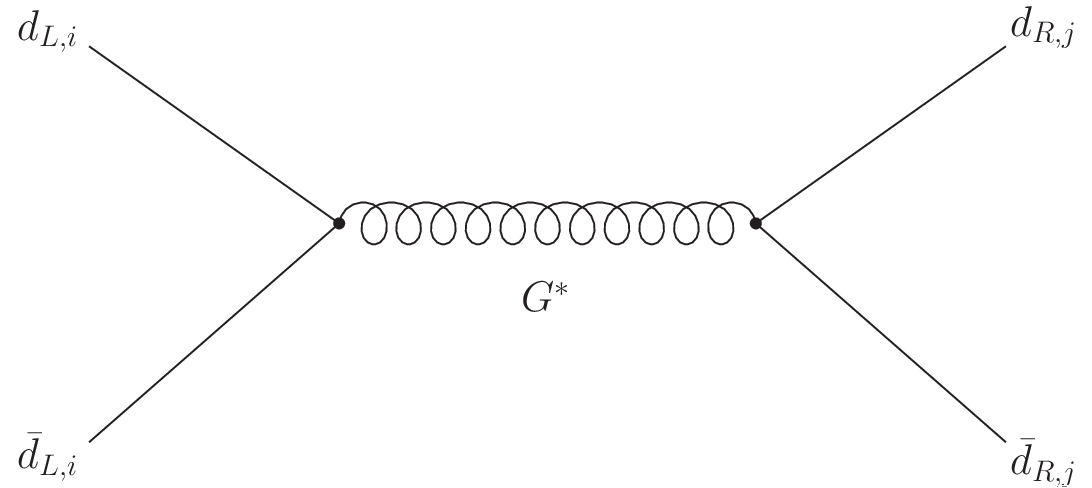, width=0.55\linewidth}
\caption{\it 
Tree-level exchange of an excited gluon, $G^*$, leading to the
four-fermion LR operator of Eq.~(\ref{4fop}).
}
\label{fcnc}
\end{center}
\end{figure}

Rather than re-discussing all the incarnations of partial compositeness in 
the phenomenology of low-energy flavor-changing processes, we will study just one 
class of examples to illustrate how the GIM mechanism of the SM is 
generalized in protecting against excessive flavor-changing neutral currents (FCNC's). 
The reader will thereby easily see how to compute other flavor-violating rare processes of interest
from the Lagrangian of Eq.~(\ref{exp fermion action}). We will focus here on 
four-fermion operators of the form
\begin{equation} \label{4fop}
{\cal L}_{FCNC} = A_{ijmn} \left( \bar{d}_{Li} \lambda_a \gamma_{\mu} d_{Lj} \right)
  \left( \bar{d}_{Rm} \lambda_a \gamma^{\mu} d_{Rn} \right)\, , 
\end{equation}
arising by integrating out the exchange of an excited gluon, $G^*$, as
in Fig.~\ref{fcnc}.
The indices $i, j, m, n = 1,2, 3$ are generational, while $\lambda_a$ are
the $SU(3)$ Gell-Mann matrices with the normalization of 
$\mathrm{Tr}(\lambda_a \lambda_b)=2\, \delta_{ab}$.

Using the Lagrangian of Eq.~(\ref{exp fermion action}), Fig.~\ref{fcnc} yields
(we consider only flavor-non-universal terms and work at leading order in $\theta_3$)
\begin{equation}
\label{fcncgauge}
A_{ijmn} \simeq
 -\frac{g_3^2 \cot^2\theta_3}{4M^2_*}\; 
 \delta_{ij} \delta_{mn}\, (\sin^2\!\varphi_{d_{L}})_i (\sin^2\!\varphi_{d_{R}})_m \, ,
\end{equation}
but it must be borne in mind that this is in the gauge-eigenstate basis prior to EWSB. 
After EWSB, FCNCs will emerge upon converting to the mass-eigenstate basis as usual. 
Denoting by $S_{d_L}$, $S_{d_R}$ the $3 \times 3$ unitary transformations of the left- and right-handed 
down quarks that diagonalize the down-quark mass matrix, we have that 
in the mass-eigenstate basis
\begin{equation}
\label{fcncmass}
A_{ijmn} \simeq -\frac{g_3^2 \cot^2\theta_3}{4M^2_*}\,
 (S_{d_L})^\dagger_{ki} (S_{d_L})_{kj}\, (\sin^2\!\varphi_{d_{L}})_k\, (\sin^2\!\varphi_{d_{R}})_l \,
 (S_{d_R})^\dagger_{lm} (S_{d_R})_{ln}   \, .
\end{equation}
We can then see how the 
generalized flavor protection works for the lighter fermions. 
Flavor-changing neutral currents 
arise via $S_{d_L}$, $S_{d_R}$, but also depends on
 the partial compositeness of the fermions, $(\sin\varphi_{d_{L}})_i, (\sin\varphi_{d_{R}})_i \neq 0$.
The lighter fermions are mostly elementary, with weak couplings to the Higgs, 
$|(\varphi_{d_{L,R}})_i| \ll 1$. Thus FCNCs are suppressed the lighter the fermions involved.

\section{Oblique corrections}
\label{oblique corrections}

\begin{figure}[t]
\begin{center}
\epsfig{figure=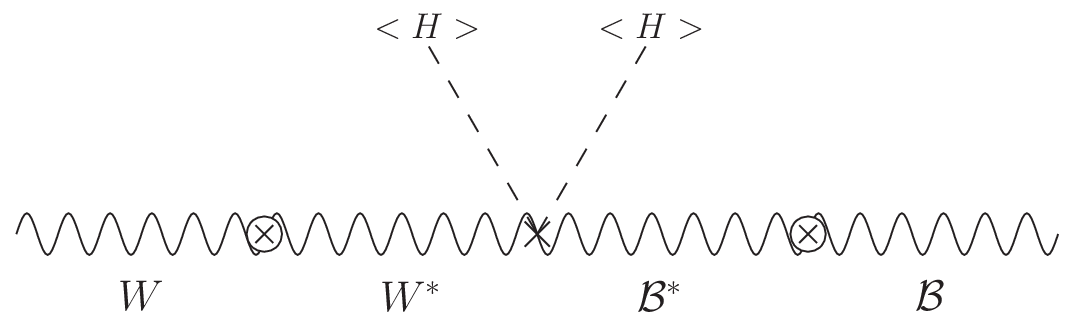,scale=0.55} 
\caption{\it Tree-level contribution to $S$.}
\label{s}
\end{center}
\end{figure}

\begin{figure}[t]
\begin{center}
\epsfig{figure=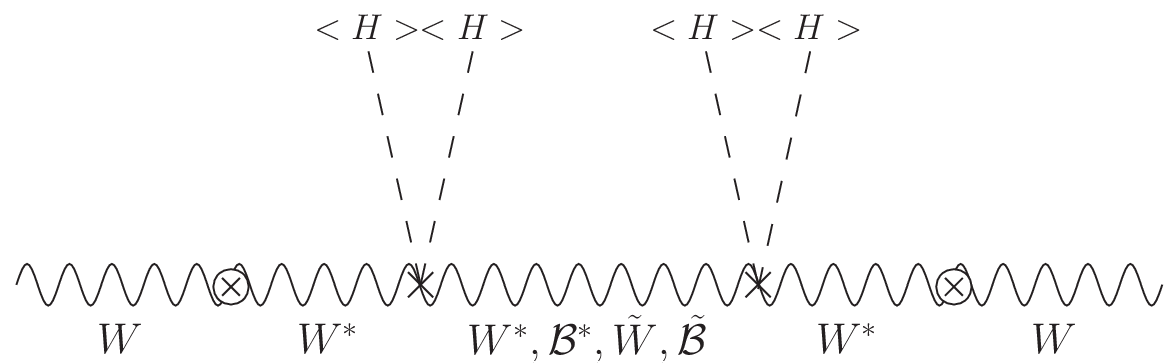,scale=0.55}
\caption{\it Tree-level contribution to $T$. The $SU(2)_L\!\times\! SU(2)_R$ invariance 
of the composite gauge sector ensures that this contribution exactly vanishes.}
\label{t}
\end{center}
\end{figure}

\begin{figure}[t]
\begin{center}
\epsfig{figure=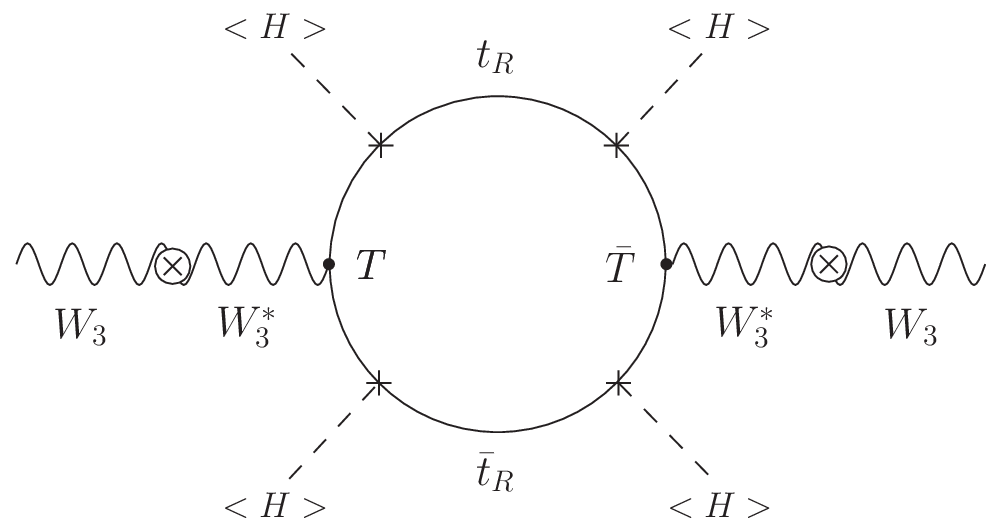,scale=0.55}
\caption{\it One-loop dominant contribution to $T$.}
\label{t loop}
\end{center}
\end{figure}

The leading oblique corrections can be parametrized by the Peskin-Takeuchi
$S$ and $T$ parameters~\cite{PT}. One useful aspect of these variables is that they 
are defined directly in terms of 
vacuum polarizations, rather than in terms of corrections to the physical SM gauge couplings.
This is done by adopting an oblique field basis, 
which is precisely the basis in which all the universal corrections
to the gauge couplings of the SM light fermions can be encoded in the electroweak 
vacuum polarizations.
Quite interestingly, the oblique basis of Peskin and Takeuchi essentially coincides with the 
elementary/composite basis, apart from highly-suppressed corrections due to the
small composite component of the physical light fermions.
This means that the new-physics contributions to these variables 
can be obtained more readily by using the Lagrangian {\it before} 
mass-diagonalization. 

The leading new-physics contribution to the $S$ parameter is given by 
the mixing diagram of Fig. \ref{s}, and yields
\begin{equation} \label{S par}
S=4 \pi v^2 \left(\frac{1}{M_{*1}^2} + \frac{1}{M_{*2}^2}  \right)\,  .
\end{equation}
We can already foresee one central feature of the 
spectrum relevant for the LHC: the composite vector mesons must be quite heavy,
$M_* \gtrsim 2.3$ TeV in order to keep $S \lesssim 0.3$, as required at the 99\% CL 
by the precision data (see for example~\cite{Barbieri:2004qk}). 

The tree-level contribution to the $T$ parameter
is given by the diagram of Fig.~\ref{t}, but the custodial symmetry of 
the composite sector implied by the $[SU(2)_L \otimes SU(2)_R]^{comp}$ 
global invariance ensures that there is a complete cancellation:
\begin{equation}\label{T par tree}
T=0  \qquad \text{at tree level} \, . 
\end{equation} 
In fact, the suppression of $T$ at the tree level by a custodial isospin symmetry 
is the motivation for extending the composite gauge group.
The custodial symmetry of the composite gauge sector is however spoiled 
at the loop level, since all composite fermions are singlets of $[SU(2)_R]^{comp}$
and their Yukawa couplings to the Higgs break explicitly the custodial symmetry.
The dominant one-loop contribution to the $T$ parameter is given by the diagram 
of Fig.~\ref{t loop}, which gives
\begin{equation}\label{T par loop}
T=\frac{2}{3}\,T_{SM}^{top}\, \left(\frac{Y_{*U 33}^2}{Y_{top}}\right)^2 \left(\frac{v}{m^Q_{33}}\right)^2 \, ,
\end{equation}
where
\begin{equation}
T_{SM}^{top} = \frac{3}{16 \pi \sin^2\theta_w \cos^2\theta_w}\frac{m_t^2}{m_Z^2} \simeq 1.2 \, .  
\end{equation}

\section{Non-universal corrections: $Z \rightarrow b_L \bar{b}_L$}
\label{sec:ztobbbar}

\begin{figure}[t]
\begin{center}
\epsfig{figure=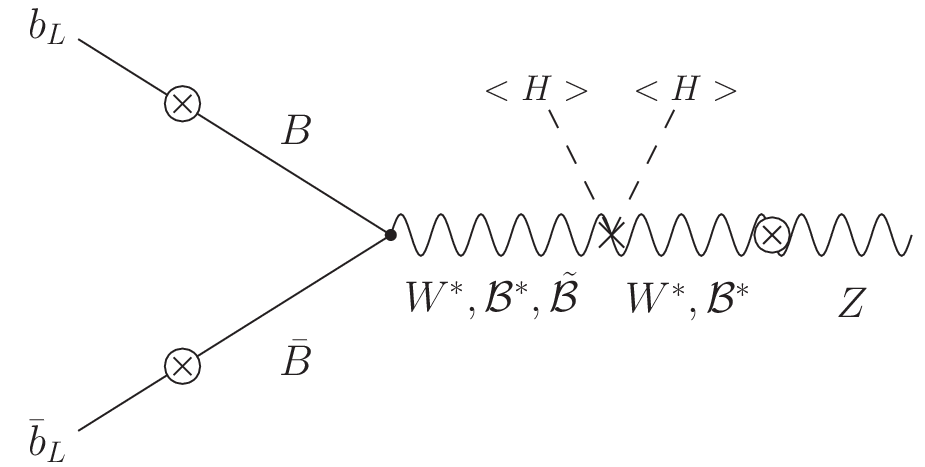,scale=0.58}
\caption{\it Tree-level contributions to $\delta g_L$ from the exchange of 
gauge excitations, in the insertion approximation. A circled
cross denotes a mass mixing.}
\label{ztobbbar}
\end{center}
\end{figure}

Non-universal effects are most important for the quarks of the third generation, since
these are the fermions with the largest mixing with the composite sector.
In particular, the strongest constraint comes from the $Zb_L \bar b_L$ coupling, which has
been quite precisely measured by LEP and SLD experiments.
We follow the usual convention and define the left-handed coupling $g_{Lb}$ so that
its SM value at tree level is $g_{Lb} = T^{3L} - Q \sin^2\theta_w$.
After EWSB $g_{Lb}$ receives a correction due to the mixing of $b_L$ with
the heavy electroweak singlet $\tilde B$. This shift is of the form
\begin{equation}
\delta g_{Lb} = \sin^2\xi \left( T^{3L}(\tilde B) - T^{3L}(b_L) \right) \, ,
\end{equation}
where $T^{3L}(\tilde B) = 0$, and the mixing angle $\xi$ between $b_L$ and $\tilde B_L$ 
can be extracted from the mass matrix (\ref{MD}). 
At leading order in $(v \sin\varphi_{b_L} Y_{*D33}/\tilde m_*^B)$ we obtain:
\begin{equation}
\xi \simeq \frac{Y_{*D33}}{\tilde m_*^B} \frac{v}{\sqrt{2}} \sin\varphi_{b_L} \cos\varphi_{b_R} \, .
\end{equation}
Another correction to $g_{Lb}$ comes from the mixing of the $Z$ with $W^{*3}$, $B^*$
and ${\cal B}$ after EWSB.
This shift is most conveniently computed adopting the elementary/composite basis,
and working for simplicity at leading order in the elementary/composite mass insertions
$(\Delta_{b_L}/m^Q)$ and $(\gel/\gc)$. The leading contribution is given by the diagram
of Fig.~\ref{ztobbbar}, which, added to the fermionic one, leads to
(for $t_R$ fully composite)
\begin{equation}\label{Zbb}
\delta g_{Lb} \simeq
  \frac{1}{2} \left(\frac{Y_{*D33}}{Y_{*U33}} \right)^2 \left(\frac{m_t}{\tilde{m}_*^B}\right)^2
+ \frac{1}{4} \left(\frac{m_t}{M_{*2}} \right)^2 \left(\frac{g_{*2}}{Y_{*U33}} \right)^2  \, .
\end{equation}

\section{Explicit gauge-invariance breaking and UV cutoff}
\label{cutoff}

While our model's defining Lagrangian has all explicit couplings and mass 
parameters with non-negative mass dimensions (naively a check of 
renormalizability), the model is still non-renormalizable because $M_*, Y_*$ 
and the mixing masses explicitly break the composite and elementary gauge 
symmetries, preserving only the subgroup of SM gauge symmetries. Therefore, 
the quantum theory becomes strongly coupled in the UV and there is a 
maximal energy, $\Lambda_{UV}$ before losing perturbative control. In this 
appendix we estimate this effective UV cutoff on the consistency of our 
model and show that it is well above the energies we wish to study. 
Of course, cutoff physics can also appear virtually in precision effects, 
but these will be suppressed relative to $M_*$-scale contributions by 
${\cal O}(M_*^2/\Lambda_{UV}^2)$.

The strongest couplings arise from the composite sector, so for the present 
exercise we focus on this sector exclusively, in particular the breaking of 
composite gauge invariance due to $M_*, Y_*$. We can always imagine that 
the composite Lagrangian is just the unitary gauge representation of a new 
Higgs mechanism (distinct from the electroweak Higgs mechanism of course). 
One can ``go back'' to the un-gauge-fixed theory by performing a 
general gauge transformation, of general form exp$(i \Pi(x). {\cal T}/F_*)$, 
on the composite Lagrangian, 
and interpreting the spacetime-dependent parameters $\Pi(x)$ as the eaten 
Goldstone bosons. For example,
\begin{equation}
\frac{M_*^2}{2} \rho_{\mu}^2 \rightarrow \frac{M_*^2}{g_*^2}
|D_{\mu} e^{i \Pi. {\cal T}/F_*}|^2,
\end{equation}
the leading term in a now explicitly non-renormalizable (gauged) chiral 
Lagrangian. The Goldstone bosons, $\Pi(x)$, are canonically normalized if we 
take their ``decay constant'', $F_*$, to satisfy
\begin{equation}
M_* = \frac{g_* F_*}{2}.
\end{equation}
We can now use naive dimensional analysis ~\cite{Cuttoff}
to estimate when the non-renormalizable interactions become non-perturbatively 
strong. The standard result for the non-renormalizable chiral Lagrangian is
\begin{equation}
\label{UVcutoff}
\Lambda_{UV} \sim 4 \pi F_* = \frac{8 \pi M_*}{g_*}\, ,  \qquad
\Lambda_{UV} \gg M_* \quad \text{for }  g_* \ll 4 \pi \, .
\end{equation}

Similarly, the un-gauge-fixed composite Yukawa coupling is of the general form
\begin{equation}
Y_* \bar\chi H \tilde{\chi} \rightarrow 
Y_* \bar\chi H e^{i \Pi. {\cal T}/F_*} \tilde{\chi}.
\end{equation}
Such a non-renormalizable coupling renormalizes itself (at two-loop order).
For the divergent loop corrections to be weaker than the tree coupling, 
the cutoff must satisfy
\begin{equation}
\frac{Y_*^3 \Lambda_{UV}^2}{(16 \pi^2)^2 F_*^2} < Y_*.
\end{equation}
The implied cutoff, $\Lambda_{UV} < 16 \pi^2 F_*/Y_*$, is subdominant to 
Eq. (\ref{UVcutoff}).

We conclude that our model is a weakly coupled non-renormalizable effective 
field theory up to an energy scale given roughly by Eq. (\ref{UVcutoff}), which 
is over $10$ TeV for the parameter range of interest.

\section{Randall-Sundrum graviton and radion excitations}

The model we have presented in this paper is the minimal one realizing 
the partial compositeness of SM fields. The original Randall-Sundrum 
warped compactification model \cite{Randall:1999ee} 
focussed instead on the presence of 
observable graviton excitations at the TeV scale, needed to realize 
partial compositeness of the ordinary massless graviton. We have not 
included the lightest graviton excitation because the massless graviton is 
itself too weakly coupled to be relevant at particle colliders and because 
the more strongly coupled graviton excitation is SM-neutral and less 
important in the face of so many SM-charged excitations, as in warped 
compactifications with ``bulk'' SM fields. The original Randall-Sundrum 
model also possessed a light SM-neutral 
``radion'' scalar, although its presence 
is dependent on details of stabilization of the 
size of the warped compactification, so again we have not included it thus far. 
Moreover, it is not needed to realize partial compositeness of just the SM. 
However, both the graviton and radion excitations have interesting 
properties and may be visible at colliders, and therefore 
deserve further consideration.
We briefly indicate how 
they are to be included in the present context.

From the compositeness viewpoint, the  
graviton and radion 
excitations are pure spin-2 and spin-0 composites respectively.
(There is entirely negligible mixing 
with the massless elementary graviton.) They are included in our model 
by modifying the composite sector and mixing terms, 
\begin{equation}
\begin{split}
{\cal L}_{composite}  
&+ {\cal L}_{mixing} \rightarrow  
 \sqrt{- G} \left\{ \tilde{\cal L}_{composite} + \tilde{\cal L}_{mixing}
 + 2 M_{Pl*}^2 R \right\}  \\
&- \frac{{\cal M}^2_*}{4} \left( 1 + \frac{\phi}{F} \right)^4  
 \left(H_{\mu \nu} H^{\mu \nu} - H_{\mu}^{\mu} H_{\nu}^{\nu} \right)  
 + \left(\frac{1}{2} + \frac{12 M_{Pl*}^2}{F^2} \right)
 \left( \partial_{\mu} \phi \right)^2 - V_{GW}(\phi) \, .
\end{split}
\end{equation} 
The first line is generally coordinate invariant with respect to a 
``metric'' field 
\begin{equation}
G_{\mu \nu} \equiv \left(\eta_{\mu \nu} + \frac{H_{\mu \nu}}{M_{Pl*}} \right) 
 \left( 1 + \frac{\phi}{F} \right)^2 , 
\end{equation}
which houses both the symmetric Lorentz-tensor graviton field, $H_{\mu \nu}$,
and the radion scalar, $\phi$. The $\tilde{\cal L}$ terms are minimally 
coupled to the metric, while the $R$ term is the Ricci scalar term 
made from $G_{\mu \nu}$, acting as a kinetic term for the graviton. 
In this field normalization, the graviton ``Planck mass'' parameter of 
several TeV controls the coupling of the graviton to itself and other 
fields. 
The second line however breaks the general coordinate invariance, by 
the graviton ``Pauli-Fierz'' mass term \cite{pf}, 
with mass ${\cal M}_* >$ TeV,   
a radion kinetic term (not coupled to the graviton), and a radion potential. 
The radion kinetic term appears peculiarly normalized because it also 
gets a contribution from the $R$ term on the first line. The
deconstructed form of the light Randall-Sundrum graviton excitation was 
proposed in Ref. \cite{mg}. From the compositeness viewpoint it relates 
\cite{schwartz}
to the strong interactions approach of Tensor Meson Dominance \cite{suzuki}.

If we turn off all the elementary sector fields as well as $V_{GW}$, 
and focus on the composite 
sector alone, the radion field is coupled consistently with it being 
the Goldstone boson of spontaneous conformal-symmetry 
breaking in the composite 
sector.  This is the composite dynamics dual role 
of the Randall-Sundrum radion \cite{AdSCFT3,dilaton}.
That is, the physical picture is that the composite sector is 
strongly coupled but conformally-invariant until this 
conformal-symmetry breaking 
at a scale $F$, below which the composites are generally massive, 
but coupled to the massless Goldstone boson, $\phi$.
The radion potential,
\begin{equation}
 V_{GW} = \frac{1}{2} m_{\phi}^2 \phi^2 + \mu_{\phi} \phi^3 + \lambda_{\phi} \phi^4 \, ,
\end{equation}
represents a weak explicit breaking of the Goldstone symmetry as 
exemplified in the Randall-Sundrum context by the well-known 
Goldberger-Wise (GW) stabilization mechanism \cite{gw}.

There is a standard 
subtlety in coupling the (composite) fermions to the graviton 
excitation, in that one cannot work in terms of a metric field, but rather 
must use a vierbein, 
\begin{equation}
E_{\mu}^{~a} = \delta_{\mu}^{~a} + \tilde{H}_{\mu}^{~a} \, , 
\end{equation}
which is related to the metric field by 
\begin{equation}
G_{\mu \nu} \equiv \eta_{ab} E_{\mu}^{~a} E_{\nu}^{~b} \, .
\end{equation}
In standard fashion, both metric and vierbein house the same physical 
graviton field, plus extra off-shell degrees of freedom, but the (more 
cumbersome) vierbein is useful for covariantizing fermion kinetic terms.

With the above modification of the composite and mixing Lagrangians, the 
reader can easily derive
the couplings of the SM and the SM-charged excitations to the radion and graviton.
Note that some of the most important couplings of the 
radion to the SM fields will appear at the loop level. For example, 
it is straightforward to check that after mass diagonalization the 
radion-gluon-gluon coupling vanishes classically, but it is generated 
via quantum loops of heavy colored excitations.

\end{document}